\documentclass[letterpaper,twocolumn,10pt]{article}
\usepackage{usenix-2020-09}

\usepackage{cite}

\usepackage{amsmath,amssymb,amsfonts}
\usepackage{xspace}
\usepackage{graphicx}
\usepackage{balance}
\usepackage{booktabs}
\usepackage{nicefrac}
\usepackage{multirow}
\usepackage{paralist}
\usepackage{footnote}
\usepackage{tcolorbox}
\usepackage{url}
\usepackage{threeparttable}
\usepackage{algorithm}
\usepackage{algorithmic}
\usepackage[binary-units=true]{siunitx}
\usepackage{pifont}
\usepackage{adjustbox}
\usepackage{tikz}

\usetikzlibrary{positioning,shapes}

\graphicspath{{src/fig/}}

\newcommand{\ie}{i.\@\,e.\@\xspace}
\newcommand{\eg}{e.\@\,g.\@\xspace}
\newcommand{\cf}{cf.\@\xspace}

\newcommand{\NEW}[2]{\textcolor{black}{#2}}

\newcommand{\verloc}{\textit{VerLoc}\xspace}

\DeclareMathOperator*{\argmin}{arg\,min}

\definecolor{analysis_blue}{rgb}{0.19, 0.55, 0.91}
\definecolor{measurement_green}{rgb}{0.0, 0.5, 0.0}

\usepackage{amsmath}

\begin{document}


\title{\verloc: Verifiable Localization in Decentralized Systems}

\author{
{\rm Katharina Kohls}\\
Radboud University Nijmegen\\
kkohls@cs.ru.nl
\and
{\rm Claudia Diaz}\\
imec-COSIC KU Leuven \\ 
Nym Technologies SA\\
claudia.diaz@esat.kuleuven.be
} 

\maketitle
\pagestyle{empty}
\begin{abstract}
We tackle the challenge of reliably determining the geo-location of nodes in decentralized networks, considering adversarial settings and without depending on any trusted landmarks. In particular, we consider active adversaries that control a subset of nodes, announce false locations and strategically manipulate measurements.
To address this problem we propose, implement and evaluate \verloc, a system that allows verifying the claimed geo-locations of network nodes in a fully decentralized manner. \verloc securely schedules roundtrip time (RTT)  measurements  between  randomly chosen pairs of nodes. Trilateration is then applied to the set of measurements to verify claimed geo-locations.
We evaluate \verloc both with simulations and in the wild using a prototype implementation integrated in the Nym network (currently run by thousands of nodes). We find that \verloc can localize nodes in the wild with a median error of \SI{60}{km}, and that in attack simulations it is capable of detecting and filtering out adversarial timing manipulations for network setups with up to \SI{20}{\percent} malicious nodes.
\end{abstract}

\section{Introduction}

Whenever network applications depend on specific locations for service nodes, they also depend on truthful location information~\cite{proxies_lie,chandrasekaran2015alidade}. As GeoIP databases are not always reliable~\cite{poese2011ip,shavitt2011geolocation,gharaibeh2017look}, active localization approaches that use timing measurements to derive geo-location have been proposed in prior work. Systems like Spotter~\cite{laki2011spotter}, Octant~\cite{wong2007octant}, or constraint-based geo-location~\cite{katz2006towards,gueye2006constraint,komosny2013can} send timing probes to targets from trusted landmarks that have a known location~\cite{dabek2004vivaldi,eriksson2010learning}. Combining the timing measurements obtained by the set of landmarks allows to narrow down the location of the target. 
While this allows for predictions up to street-level granularity~\cite{wang2011towards,chen2016landmark}, landmark-based systems rely on a trusted setup. Spotter and Co. depend on accurate ground truth information for landmark locations, as well as on honest accurate reporting of timing measurements by the landmarks. Such systems are neither robust to malicious landmarks that lie about their location or obtained measurements, nor to malicious targets that strategically manipulate timing measurements by, \eg, delaying responses to certain timing probes. This makes these solutions inadequate for decentralized settings that may be subject to adversarial conditions.


A scheme that allows to verify geo-location in networks in a fully decentralized manner -- without relying on trusted landmarks or measurements -- can be useful in a variety of scenarios. Here we highlight two use cases. First, \textit{overlay anonymous communication networks} such as Tor~\footnote{\url{https://www.torproject.org}} and Nym~\footnote{\url{https://nymtech.net}} route user connections through relays in multiple jurisdictions to protect against adversaries who have monitoring and coercion powers within a zone of adversarial control. Location diversity strengthens security, as it becomes harder to monitor, compromise or censor the full network~\cite{SoK-decentralization}.
Ensuring location diversity when routing a connection requires reliable geo-location information. However, prior work demonstrates that available solutions sometimes result in incorrect location information, \eg, \num{194} out of \num{6042} Tor relays were found to be in a different country than the one indicated by their GeoIP entry~\cite{kohls19multi}. We can thus see that misleading location information is not only a theoretical possibility, and may even be actively used to obfuscate the whereabouts of network nodes. In addition to supporting geographic diversity, publicly verifiable node locations enable other functionalities dependent on accurate location data, such as location-aware anonymous routing policies that reduce end-to-end latency by favouring routes that travel a smaller distance~\cite{akhoondi2012lastor,claps}. 

Second, location diversity is also important for resilience purposes in \textit{peer-to-peer networks that jointly maintain a blockchain}, such as Bitcoin~\footnote{\url{https://bitcoin.org/}}. The concentration of network servers in certain geo-locations makes the network vulnerable to regional events, including natural disasters~\cite{coindesk} as well as politically motivated interventions~\cite{china_destroys_bitcoin}. A method to reliably verify peer locations in a fully decentralized manner would enable such permissionless networks to incentivize location diversification, while ensuring that malicious peers cannot take advantage by faking their location. In particular, node locations can serve as one of the variables in the delegation criteria in systems based on delegated proof of stake~\cite{ouroboros}.

These functionalities are compelling not just from an academic standpoint. The Nym network~\cite{nym-whitepaper} has already integrated and deployed a prototype implementation of \verloc that Nym mix nodes run twice a day. Nym wants to verify node locations to be able to enable in the future: (1) routing policy constraints to ensure routes traverse multiple jurisdictions and better protect from nation-state adversaries, (2) lower-latency location-aware routing, and (3) incentives for global location diversification via rewards and delegation of stake (\eg premium reward rates for nodes located in geographical areas with lower node density). At the time of writing \num{3460} nodes have upgraded to the \verloc-enabled version. We take advantage of this experimental prototype implementation to collect measurements and validate \verloc's performance in the wild.

\noindent \textbf{Contribution.}
\verloc tackles the challenge of verifying geo-locations in a fully decentralized network without trusted authorities or landmarks, where up to \SI{20}{\percent} of the network may be actively malicious. To do so, \verloc uses a novel timing-based verification algorithm that is robust to network noise and can withstand strategic adversarial manipulation. \verloc securely schedules Round Trip Time (RTT) measurements so that the adversary cannot influence randomized assignments. Based on pairwise RTT measurements, \verloc applies trilateration to estimate the geo-location of nodes and verify their claimed whereabouts. \verloc uses a broadcast channel to enable all nodes to share the information needed to verify all geo-locations. The measurement overhead involves sending a few thousand pings, and it remains constant as the network grows, while the data processing and storage overhead grows linearly with the number of nodes (by \SI{200}{\byte} per node). 

As preliminary step, we conduct an empirical study to derive a realistic network propagation model that accounts for the effects of dynamic routing, congestion, and other naturally occurring noise. This propagation model enables \verloc to convert measured times into geographical distances while accounting for the effects of noise on the confidence intervals. 

We first assess \verloc's performance baseline in the absence of active adversaries, \ie, considering that all network nodes honestly report their location and measurements. 
We conduct extensive simulations to evaluate the performance of \verloc under different conditions and understand the effects of parameters that affect the accuracy of the results. We present results for a challenging deployment scenario and show that even in sub-optimal conditions \verloc is able to localize nodes within a median error range of \SI{103}{\km} and verify with accuracy \SI{92}{\percent} the country where a node is located. Repeating the experiments in the wild provides even better results, with a median localization error of \SI{60}{\km}. We then evaluate \verloc against an adversary that controls a subset of nodes, considering that malicious nodes may lie about their location, report fake timing measurements, and even manipulate pairwise measurements by delaying responses. We introduce a \emph{confidence score} that qualifies location verification decisions and show that the score is effective for accurately distinguishing between true (honest) and false (adversarial) reported locations. We find that adversaries need to control more than \SI{20}{\percent} of nodes to begin to degrade the location verification accuracy for honest nodes, and more than \SI{30}{\percent} to trick \verloc into accepting fake locations.





\section{Preliminaries}\label{sec:preliminaries}

\subsection{Problem Statement}\label{subsec:problem_statement}

We consider a network of servers, which we refer to as \emph{nodes}, that are geographically disperse and work together to enable a service, \eg, the relays that constitute the Tor network or the peers that are part of the Bitcoin network. Note that the network may provide services to \emph{end clients} that do not publicize their contact and location information or take part in the verification process. We consider that only the nodes that form the network infrastructure participate in the \verloc protocols. 

Network nodes announce their geographical location and conduct a limited number of pairwise RTT measurements that they also broadcast. Based on the set of claimed locations and pairwise measurements, \verloc allows everyone to verify all the claimed locations in a fully decentralized manner. \verloc is designed to function in an adversarial environment where a subset of malicious nodes coordinate to claim false locations. The goal of \verloc is to distinguish between true and false locations with high accuracy, even in the face of random network noise and active adversaries \NEW{6}{that control a subset of colluding nodes. Considering an adversary that claims false node locations and manipulates reported  measurements, \verloc's main security goals can be formulated as follows: first, an adversarial node $n_a$ at geo-location $l_a$ cannot successfully claim being at a distant location $l_a'$ (e.g., that is in a different country); and second, honest nodes that correctly follow the protocols and report truthful information are successful in verifying their geo-location.}

\noindent \textbf{Network model.} We model the network as a set of $N$ nodes $n_i$ with $i=1..N$. Each node $n_i$ broadcasts a descriptor with its public key $pk_i$, its network address $IP_i$, and its geographical location $loc_i$ (ideally with an error of less than \SI{10}{\km}), specified by latitude and longitude. This information is periodically broadcast and nodes are `committed' to their keys, address and location until the next update. We also assume that it is possible for any two nodes $n_i$ and $n_j$ to communicate directly via the Internet, \ie, the network graph is fully connected.  

\noindent \textbf{Broadcast channel.} We assume that each node knows the full set of $N-1$ other nodes, with their \textit{node descriptors} (including $pk_i$, $IP_i$, and $loc_i$). This may be achieved in different ways. In anonymous communication networks this information is typically updated in a consensus document published every hour or few hours, while in peer-to-peer networks maintaining blockchains the information may be continuously updated via gossip protocols. The timing measurements collected by nodes, which are needed to verify locations, must also be broadcast to ensure public verifiability of results. \verloc thus requires a broadcast channel. We propose using a blockchain that acts as a public, append-only log maintained in a decentralized manner by the nodes. Note however that this blockchain can be replaced by any technology that provides secure broadcast with Byzantine fault-tolerance~\cite{byzantine-generals}. 
\NEW{1}{The key security features required by \verloc from the channel are integrity and availability, i.e., that once uploaded, information is publicly available to all participants and cannot be altered. Node descriptors must be digitally signed to ensure that they have been generated by the node associated to the descriptor's public key and have not been altered by others. In turn, the list of descriptors of the nodes that constitute the network for a period of time is jointly signed by the entities maintaining the broadcast channel. We assume all participants can authenticate the channel without being tricked into believing that a separate channel (controlled by the adversary) is the authentic one containing \verloc's information. Furthermore we assume that it is not possible for the adversary to censor participant's read or write access to the channel, i.e., all nodes are able to broadcast their descriptor and measurements, and to read the descriptors and measurements of all the other nodes.}

\noindent \textbf{Epochs.} We consider that nodes commit to being available at a location for a finite amount of time. The network is updated periodically, \eg every few hours, with the \emph{epoch} length being dependent on the expected churn in the network. Before the start of an \emph{epoch}, nodes broadcast their updated keys, addresses and locations. The \verloc protocols run during the epoch and produce results that enable identifying malicious nodes and possibly excluding them from the next epoch. 

\noindent \textbf{Timing information.} \verloc relies on \emph{timing information} to estimate locations, using the relation between measurable transmission times and geographical distances between nodes, which is bound by the speed of light. More precisely, the transmission time defines the area that can be reached within that amount of time -- with larger distances being impossible to reach, as that would imply transmissions speeds that are faster than light. 

\noindent \textbf{Timing probes.} Nodes in the network probe a subset of other nodes and measure the round trip time (RTT), \ie, the time elapsed between sending a request and receiving a response. These pairwise measurements do not require any central authority and can be conducted using existing protocols such as the Internet Control Message Protocol (ICMP).

\noindent \textbf{Trilateration.} Combining transmission times measured from reference points situated in different directions allows to narrow down the location of a node. As all nodes measure various other nodes in \verloc, there is redundancy in the overall set of network measurements. This redundancy allows to detect inconsistencies created by the malicious activity of adversarial nodes, as well as distortions introduced by exceptionally bad network conditions. 

\noindent \textbf{Inference of geographical coordinates.} Based on all the claimed node locations and reported timing measurements, \verloc estimates the most likely geographical coordinates (latitude, longitude) of each node and checks the distance to the node's claimed location. 

\noindent \textbf{Most likely geographical zone.} We consider that space may be divided into countries, regions, zones, or any other territorial division, with each location (and therefore each node) belonging to precisely one zone. In addition to the most likely geographical coordinates, \verloc computes the probability that a node is located within a zone.  

\noindent \textbf{Confidence scores.} Finally, \verloc compares the set of empirically measured propagation times with the times one would expect given the locations claimed by all nodes and the propagation model. This is used to define a \emph{confidence score} that expresses the discrepancy between expected and measured times. A low confidence score is indicative that a node localization result (coordinates as well as zone) may be wrong and possibly malicious.

\subsection{Information Propagation Model}
\label{subsec:prelim:propagation}

Considering propagation \emph{speed}, the transmission \emph{time} between two nodes is proportional to the \emph{distance} between them. However, network effects introduce noise and variance in the transmission speed, and consequently error when estimating nodes' locations from measured times. To provide real-world capabilities, \verloc must be robust to realistic levels of noise and account for the actual speed function in the underlying Internet. We conduct experiments where we measure the timings of Internet transmissions between servers placed in different locations around the world. From these experiments we distill a realistic propagation model that is shown in Figure~\ref{fig:noise_overhead}. The steps we took to derive this model are explained in detail in Appendix~\ref{details-propagation-model}. Note that the propagation model is pre-computed \textit{once} and provided as a component of \verloc. While a better model can be created with a larger number of measurements, once it is obtained it does not need to be updated until the Internet infrastructure undergoes a significant enough update to change its overall propagation characteristics.  

\begin{figure}[t]
	\centering
	\includegraphics[width=.9\columnwidth]{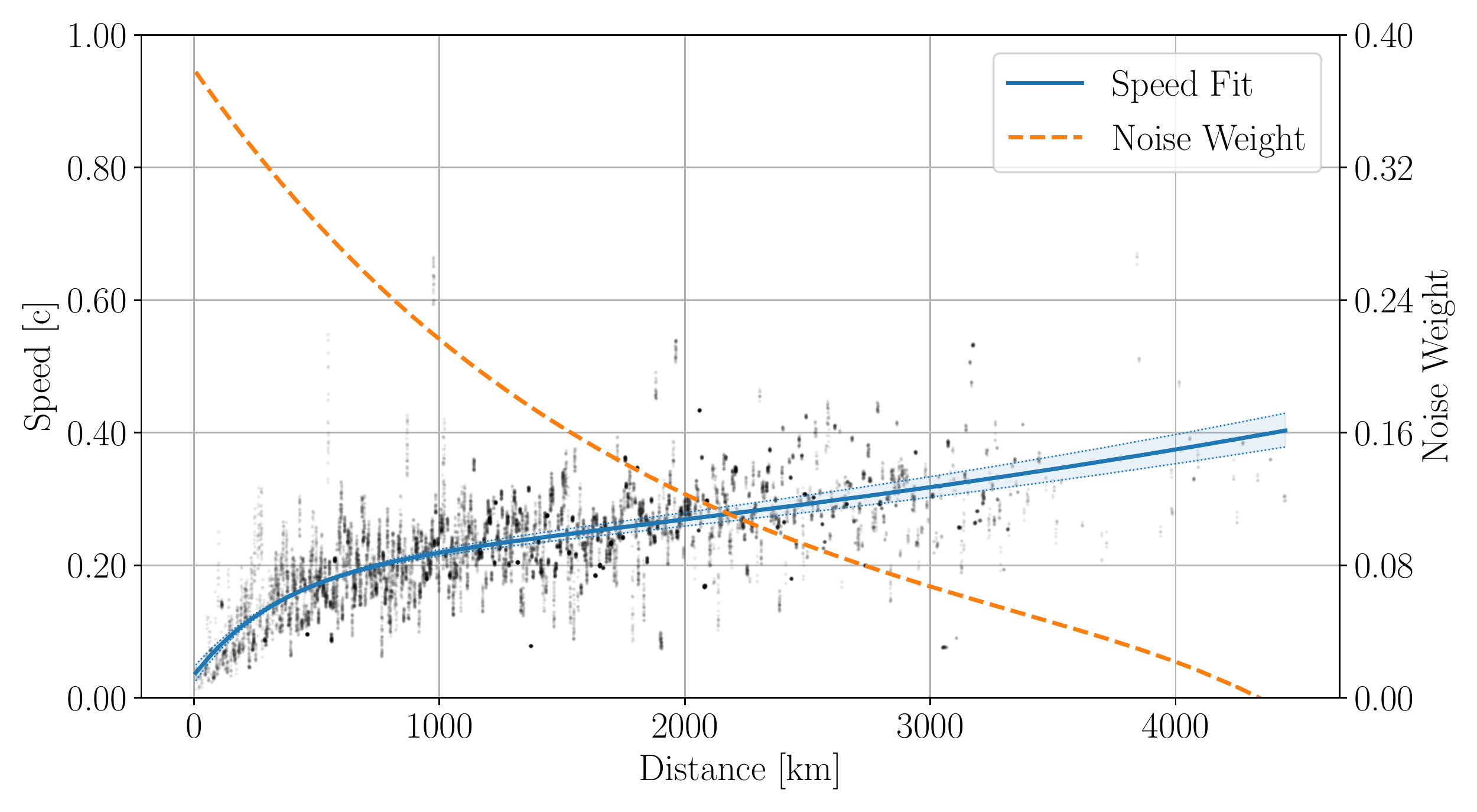}
	\caption{Propagation Speed Model. The scattered points show the individual speeds of ICMP traffic RTTs; the \emph{blue} is the speed fit; the \emph{orange} line summarizes the impact of background noise.}  
    \label{fig:noise_overhead}
\end{figure}

We find that transmission speed is, in practice, more complex than a simple constant due to \textit{background noise} caused by varying transmission medium characteristics, asymmetric and dynamic routing, congestion, and a host of other effects. As shown in Figure~\ref{fig:noise_overhead}, propagation speed and background noise depend on end-to-end distance. By applying a fitting function, we can model the scattered and noisy transmission speeds as an estimate $f(x)$ that provides us the most likely propagation speed for a given distance $x$. The speed function $f(x)$ (and its inverse $f^{-1}(t)$, which converts times to distances) is used to estimate node locations (\S\ref{subsubsec:system_estimate_locations}) and compute confidence scores (\S\ref{subsubsec:confidence_scores}). 
Furthermore, the variance of observed times for similar distances allows us to capture background noise characteristics. We find that noise is lower for longer transmissions. To account for this effect, we compute a noise weight that we later use for localization.


Our propagation model is based on \emph{real-world data} and therefore captures the actual transmission characteristics of the Internet. We base our simulation experiments on this model, and thus our sampled transmission times incorporate congestion latency and variance due to dynamic routing that is characteristic of the Internet. 

\section{System Concept}\label{sec:system_concept}
In this section we introduce the architecture and core system components of \verloc. We explain how the \emph{measurement component} schedules the random selection of references and the symmetric timing measurements, and how the \emph{localization and verification component} analyzes the measured timings to produce location verification results.

\subsection{System Components}\label{subsec:system_components}

As illustrated in Fig.~\ref{fig:process}, \verloc consists of a \textbf{measurement component} (green) that outputs timing measurements for selected pairs of nodes, and a \textbf{location verification component} (blue) that analyzes those timings to estimate the locations of network nodes. In addition, the nodes collaboratively maintain a publicly accessible \textbf{broadcast channel}~\cite{byzantine-generals}, allowing all nodes to broadcast their information and access the information broadcast by others. This makes the verification process fully \emph{decentralized}, as everyone can locally compute localization results based on broadcast data. The blockchain stores the public network parameters, the node descriptors (public key, IP address, claimed location), a per-epoch random beacon, and the reported RTT measurements. 

\pgfdeclarelayer{background}
\pgfsetlayers{background,main}
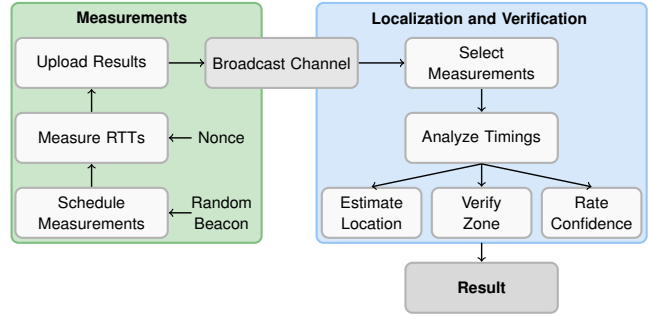
\begin{figure}[tb]
	\centering
\begin{adjustbox}{width=\columnwidth}
	
	\begin{tikzpicture}[
		squarednode/.style={rectangle, draw=gray!60, fill=gray!5, very thick, rounded corners, minimum size=10mm, text width=8em, text centered},
		smallnode/.style={rectangle, draw=gray!60, fill=gray!5, very thick, rounded corners, minimum size=10mm, text width=5em, text centered},
		textnode/.style={minimum size=10mm, text width=9em, text centered},
		font=\sffamily
		]
		\node[squarednode] (m1) at (-5.8,0) {Upload Results};
		\node[squarednode] (m2) at (-5.8,-1.5) {Measure RTTs};
		\node[squarednode] (m3) at (-5.8,-3) {Schedule\\Measurements};

		\node[textnode]    (t1) at (-3.2,-1.5) {Nonce};
		\node[textnode]    (t2) at (-3.2,-3)   {Random\\Beacon};

		\node[squarednode, draw=gray!60, fill=gray!20] (bc) at ( -2,0) {Broadcast Channel};

		\node[squarednode] (a1) at ( 2,0) {Select\\Measurements};
		\node[squarednode] (a2) at ( 2,-1.5) {Analyze Timings};
		\node[smallnode] (a3) at ( -0.2,-3) {Estimate\\Location};
		\node[smallnode] (a4) at ( 2,-3) {Verify\\Zone};
		\node[smallnode] (a5) at ( 4.2,-3) {Rate\\Confidence};

		\node[squarednode, draw=gray!60, fill=gray!30] (r) at ( 2,-4.5) {\textbf{Result}};
		
		\draw[->, thick] (m3.north) -- (m2.south);
		\draw[->, thick] (m2.north) -- (m1.south);

		\draw[->, thick] (-3.8,-1.5) -- (m2.east);
		\draw[->, thick] (-3.8,-3) -- (m3.east);

		\draw[->, thick] (m1.east) -- (bc.west);
		\draw[->, thick] (bc.east) -- (a1.west);

		\draw[->, thick] (a1.south) -- (a2.north);
		\draw[->, thick] (a2.south) -- (a3.north);
		\draw[->, thick] (a2.south) -- (a4.north);
		\draw[->, thick] (a2.south) -- (a5.north);

		\draw[->, thick] (2,-3.6) -- (r.north);
	
		\begin{pgfonlayer}{background}
		  \path[fill=measurement_green!20,rounded corners, draw=measurement_green!50, very thick] (-7.4,1.2) rectangle (-2.4,-3.6) ;
		  \path[fill=analysis_blue!20,rounded corners, draw=analysis_blue!50, very thick] (-1.3,1.2) rectangle (5.3,-3.6) ;
		
		  \node[text width=5cm] at (-3.6,0.9) {\textbf{Measurements}};
		  \node[text width=5cm] at (2.3,0.9) {\textbf{Localization and Verification}};  
		\end{pgfonlayer}
	\end{tikzpicture}
	\end{adjustbox}
	\caption{Complete measurement and analysis process. The green part includes steps involved in conducting RTT measurements, the blue part depicts the location estimation.}
    \label{fig:process}
\end{figure}


\subsection{Measurement component}\label{subsec:formal_measurements}
The measurement component involves three tasks that are executed by all network nodes: 

\begin{enumerate}
	\item \textbf{Schedule Measurements.} 
	\NEW{4}{In each epoch, the network derives nodes' reference sets, which determine the schedule of pairwise measurements,  using as seed a random beacon~\cite{random-beacon}  (\S\ref{subsubsec:scheduling_measurements}).}
	\item \textbf{Perform Measurements.} Nodes conduct pairwise measurements according to the scheduled reference sets by sending timing probes and recording the observed RTT (\S\ref{subsubsec:conducting_measurements}).
	\item \textbf{Upload Measurements.} Nodes broadcast the minimum measured RTT for each node in their reference set (\S\ref{subsubsec:uploading_measurements}).
\end{enumerate}

\subsubsection{Schedule Measurements}\label{subsubsec:scheduling_measurements}
\verloc is designed to function in a fully decentralized fashion, without relying on any trusted authorities or third parties, and with all nodes performing the same tasks. To provide robustness, it is crucial that the reference sets are chosen in a way that cannot be biased by an adversary. Otherwise, the adversary may manipulate and exploit reference sets, \eg, selecting adversarial reference sets to successfully verify false locations and reject true locations. To prevent this, \verloc assigns reference sets pseudorandomly. 

\verloc's reference set construction algorithm scales to arbitrarily large networks while maintaining a \textit{constant} (rather than quadratic) complexity in terms of the number of measurements conducted per node. As shown in Sect.~\ref{subsubsec:pb_num_refs}, the localization accuracy of \verloc increases with the number of references per node, but the improvement has diminishing returns and, after a certain point, additional references consume resources without significantly improving performance.

\noindent We empirically determine that the best tradeoff is between \num{40} and \num{80} references per node, and use those values in our experiments. 

\noindent \textbf{Reference Set Construction.}
Reference sets are derived from the node's public keys $pk_i$ and a random beacon $x$ that is jointly computed by nodes and published in the blockchain once all public keys have been committed. Alternatively, the random beacon may be obtained from an external source of randomness~\cite{kokoris2018omniledger}, \NEW{4}{e.g., the hash of the first bitcoin block published after the start of the epoch. The beacon must only become available once all nodes have committed to their public keys and are ready for a new run of the \verloc protocol. The key security requirements are that the same $x$ is available to all nodes, and that the adversary can neither determine the value of $x$, nor predict it before committing to its node public keys. Given $x$,} everyone can derive a random hash $h_i$ for node $i$ as: $h_i = H(x||pk_i)$, where $H()$ is a hash function~\cite{keccak}. 

We denote as $R_i$ the \emph{reference set} of node $n_i$. Given $h_i$, nodes follow Algorithm~\ref{alg1} to derive an initial set of $t$ references to be included in $R_i$. 
\begin{algorithm}
	\caption{Derive initial reference set $R_i$ for node $n_i$}
	\begin{algorithmic} 
	\label{alg1}
	\STATE $R_i := \emptyset$
	\STATE $y := h_i$
	\WHILE{$|R_i| < t$}
	\STATE $r := y \mod N$;
		\IF {$r \notin R_i$ and $r \neq n_i$}
			\STATE $R_i.add(r)$
		\ENDIF
		\STATE $y := Hash(y)$
	\ENDWHILE
	\RETURN $R_i$
\end{algorithmic}
\label{alg:base_references}
\end{algorithm}
These $t$ references are only a part of a node's reference set $R_i$. In \verloc, references are \emph{symmetric}, meaning that $n_j \in R_i \iff n_i \in R_j$.  Note that $n_j$ can verify that $n_i$ correctly selected it, using $h_i$ and Algorithm~\ref{alg:base_references}. Nodes complete their reference set $R_i$ following Algorithm~\ref{alg2}.  

\begin{algorithm}
	\caption{Complete reference set $R_i$}
	\begin{algorithmic} 
	\label{alg2}	
	\FOR{$j=1..N$}
		\IF {$n_j \notin R_i$ and $n_i \in R_j$}
			\STATE $R_i.add(n_j)$
		\ENDIF
	\ENDFOR
	\RETURN $R_i$
\end{algorithmic}
\label{alg:extension_references}
\end{algorithm}

With this algorithm, the complete reference set $R_i$ will have a variable size depending on the instance, as it is the sum of two components: the $t$ references (constant number) derived from $h_i$, and an additional $t'$ references (variable number) that are derived from all the other $h_j, i \neq j$. Note that the $t$ nodes that are \emph{already} included in $R_i$ do not add any new reference to $R_i$ even if $n_i \in R_j$. Given the network size $N$ and the number of references $t$, a new node $n_j \notin R_i$ is added to $R_i$ with probability $\frac{t}{N}$, and this applies to all the $N-t$ nodes that are not in the initial $R_i$. The probability of $t'$ taking a certain value $k$ thus follows a binomial distribution:

\begin{equation}\label{eq-R-t}
	Pr\left[t'=k;N-t, \frac{t}{N}\right] = \binom{N-t}{k}\cdot \left(\frac{t}{N}\right)^{k} \cdot \left(1-\frac{t}{N}\right)^{N-t-k}
\end{equation}

We select the parameter $t$ to ensure that with overwhelming probability all $N$ nodes have sufficient $t+t'$ references. 

\label{subsec:discussion:optimal_atk}
\noindent \textbf{Symmetric Measurements.} We use symmetric measurements for two reasons. First, they help leveling out noise and effects of asymmetric routing, as the timing $RTT(n_i \rightarrow n_j)$ might differ from $RTT(n_j \rightarrow n_i)$. Moreover, burst noise in one direction does not necessarily occur in the other direction and thus averaging both directions improves the overall robustness to noise. Second, averaging the times measured in both directions improves the robustness of \verloc not just towards random noise, but also active attacks. In settings where of two nodes involved in the measurement one is honest and the other malicious, the adversary could try to report a very short transmission time to manipulate (\emph{speed up}) the average transmission time. Similarly, the average could be \emph{slowed down} to a desired number by the adversary reporting a very large time.  
\emph{However, these attempts at manipulating the average are easily detectable by the confidence score (\S\ref{subsubsec:confidence_scores}).} Speeding up the average transmission means that the adversary must contribute a measurement that is significantly \emph{too fast}. Overly fast transmissions (faster than $\nicefrac{2}{3}\cdot c$) violate the upper speed bound and decrease the confidence score. Reporting a measurement that is significantly \emph{too slow} will violate the lower speed bound and also result in marking the measurement as unreliable. This lowers the confidence score for both the target and the adversary nodes, meaning that adversaries that lower the confidence score of honest nodes will cause their \textit{own} confidence score to diminish in equal measure.

\subsubsection{Conduct Measurements}\label{subsubsec:conducting_measurements} 

Given a set of references $R_i$ for node $n_i$, \verloc conducts pairwise measurements $RTT(n_i \leftarrow r_j)$ and $RTT(n_i \rightarrow r_j)$ between $n_i$ and all its reference nodes $r_j \in R_i$. A node $n_i$ conducts a measurement $RTT(n_i \rightarrow r_j)$ by sending timing probes to $r_j$, to which $r_j$ responds as fast as possible, and recording the round trip time (RTT) of the responses. In their simplest form, timing probes can be implemented, \eg, as ICMP echo requests.

\noindent \textbf{Freshness.}
To guarantee freshness, the node that initiates the timing probe includes a locally generated random nonce. When using ICMP, it is possible to encode the nonce in the variable-length data field. The responding node has to copy this nonce in the response. This prevents adversaries from \emph{speeding up} measurements. More precisely, it is always possible for a node to hold back the response to incoming probes, increasing the RTT and faking a longer transmission distance (\S\ref{subsec:atk_timing_manipulations}). By including an unpredictable nonce, an adversarial node cannot respond to an incoming probe \emph{before} it arrives, which eliminates the capability to fake a shorter distance. The nonce can be hashed with a previously established shared secret to protect against man-in-the-middle adversaries (\S\ref{subsec:discussion_attack_variants}). 


\noindent \textbf{Multiple Probes and Minimum RTT.}
The number of timing probes sent between a pair of nodes to conduct one measurement strikes a tradeoff between measurement accuracy and overhead. Increasing the number of probes allows to better overcome high-frequency, high-delta noise at the cost of sending more messages and taking longer to complete the protocol. As with measurements taken to infer the propagation model, nodes take the \emph{minimum} RTT of all the timing probes exchanged with another node as the \textit{least noisy} value. In our experiments we use series of \num{200} probes and extract the minimum value for each $RTT(n_i \rightarrow n_j)$.

\subsubsection{Upload Measurements}\label{subsubsec:uploading_measurements}
Nodes broadcast the minimum measured RTT with each of their references. These measurements are the input for the localization and verification steps.

\subsection{Localization and Verification}\label{subsec:formal_verification}

The analysis steps of \verloc are executed by all nodes locally using the reported RTT measurements as input. \verloc outputs three types of results for each node $n_i$: (1) an estimate of its location coordinates $\widehat{loc_i}$, (2) a binary verification decision for the node's claimed localization zone $z_i$, and (3) a score $c_i$ that indicates \verloc's confidence in the previous two results. 

\subsubsection{Estimate Location Coordinates}\label{subsubsec:system_estimate_locations}

\begin{figure}[t]
	\centering
	\includegraphics[width=.8\columnwidth]{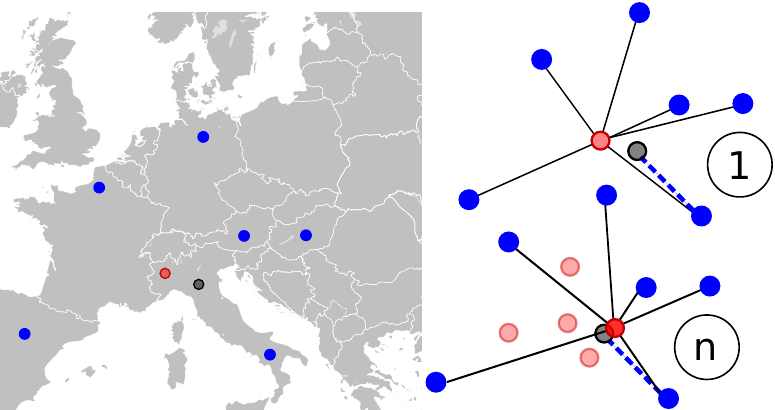}
	\caption{Node localization process (\S\ref{subsubsec:system_estimate_locations}). At the start of the localization process, we collect claimed locations and timing measurements from each of the references (blue circles) of a target node (black circle). We guess an initial location (red circle) for the target and then apply the iterative optimization. In the first step \ding{192}, we take the distance between each reference's claimed location and the guessed location (black lines). We compare this with the distance that corresponds to the \emph{measured} RTT (blue dashed line) and evaluate the discrepancy between both values. In the following steps \textcircled{n}, we use a gradient descent to adjust the guessed location (red circle) until we find the solution with the least discrepancy.}
    \label{fig:trilat_process}
\end{figure}

\verloc uses a gradient descent algorithm to estimate the geographical location of $n_i$. This optimization model is iterative and repeats three steps (illustrated in Figure~\ref{fig:trilat_process}) until it finds an estimated location $\widehat{loc_i}$ with minimum error. The process is based on the principle of trilateration~\cite{jansen2018crowd,1498470,kohls19multi} and defined as follows.

\noindent \textbf{Define Candidate Location.}
In the first step, we compute the pairwise great circle distances $dist$ between the claimed locations of nodes in the reference set $r_j \in R_i$ and a \emph{possible} candidate location $\widehat{loc_i}$ for node $n_i$. Note that we assume that most nodes are non-adversarial and claim a \emph{correct} location. The initial $\widehat{loc_i}$ is an educated guess for the location of $n_i$ that is adjusted throughout the optimization steps to find the best result. The resulting vector $\overrightarrow{dist}$, of size $R=|R_i|$, contains the distances between all references $r_j$ and candidate $\widehat{loc_i}$.

\noindent \textbf{Estimate Distances.}
We average measurements in both directions to estimate the distance between $n_i$ and $r_j$ (\S\ref{subsec:discussion:optimal_atk}). 
\begin{equation}
\label{eq:simmetricRTT}
RTT(n_i\leftrightarrow r_j) = \frac{RTT(n_i\rightarrow r_j) + RTT(n_i \leftarrow r_j)}{2}
\end{equation}
$RTT(n_i\rightarrow r_j)$ and $RTT(n_i \leftarrow r_j)$ are the \emph{minimum} RTTs measured in each direction. 
As described in Section~\ref{subsec:prelim:propagation}, we can translate transmission time into distance by applying the inverse speed function $f^{-1}(t)$. This results in a second distance vector $\overrightarrow{dist_{RTT}}$ that contains $f^{-1}(RTT(n_i\leftrightarrow r_j))$ for $r_j \in R_i$


\noindent \textbf{Apply Error Function.}
In the third step, we compare the candidate distances $\overrightarrow{dist}$ with the distances $\overrightarrow{dist_{RTT}}$ derived from the measured RTTs. The delta between both vectors $\Delta(\overrightarrow{dist},\overrightarrow{dist_{RTT}})$ expresses the discrepancy between the reported RTTs and the candidate location $\widehat{loc_i}$. We apply the root-mean-square error (RMSE) as an error function to evaluate how consistent $\widehat{loc_i}$ is with the reported measurements. We use an iterative optimization to minimize the RMSE for possible values of $\widehat{loc_i}$:
\begin{equation}
	\label{eqn:solver}
	\argmin_{\widehat{loc_i}} \sqrt{\frac{\sum_{j=1}^R (\Delta_j \cdot \omega_j)^2}{R}}
\end{equation}
$\Delta_j = |dist(\widehat{loc_i}, r_j)-f^{-1}(RTT(n_i\leftrightarrow r_j))|$ is the difference between the candidate and the measured distances for reference $r_j$, $R$ is the number of references in $R_i$, and $\omega_j$ is a distance-dependent \emph{weighting} factor that accounts for noise effects. We derive $\omega_j$ empirically as part of the propagation model introduced in Section~\ref{subsec:prelim:propagation}.

\noindent \textbf{End Result.} At the end of this process, we obtain a location estimate $\widehat{loc_i}$ that best fits the reported RTT measurements between $n_i$ and its references.

\subsubsection{Verify Zone}\label{subsubsec:location_verification}

\begin{figure*}[t]
	\centering
	\includegraphics[width=.8\textwidth]{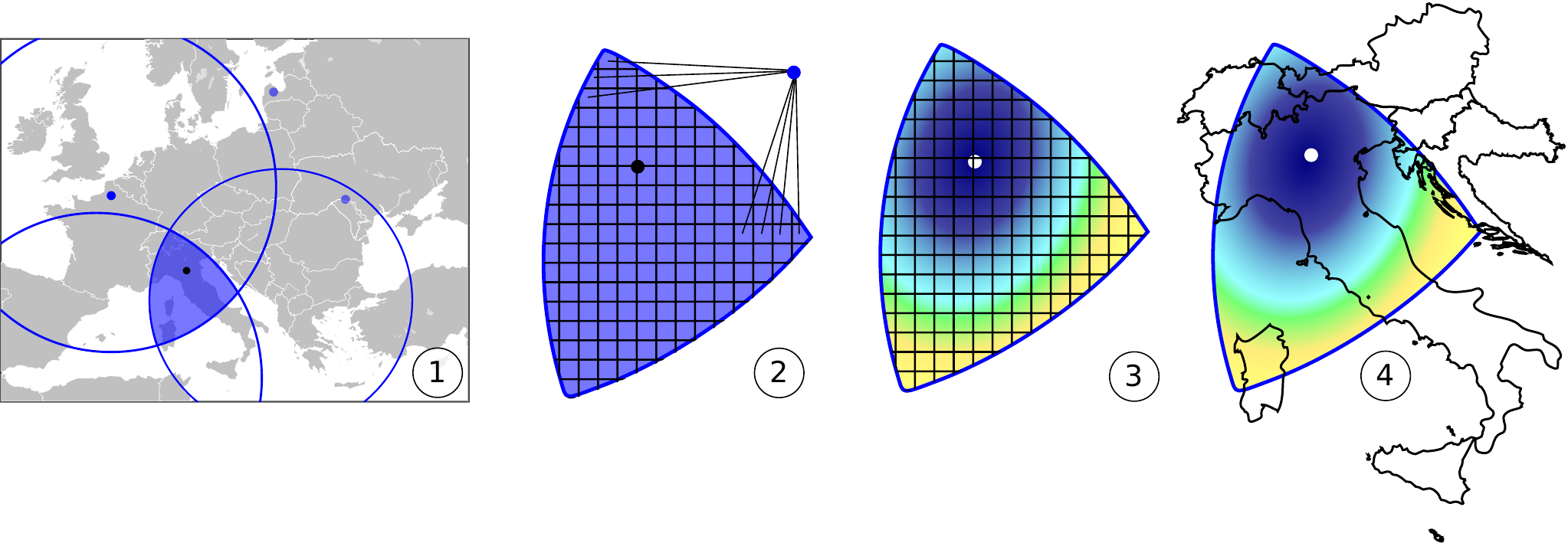}
	\caption{Zone Verification Process (\S\ref{subsubsec:location_verification}). In the first step \ding{192}, we generate an intersection area containing all possible locations for the node. In the second step \ding{193}, we apply a grid to the intersection and compute the distances between each point in the grid (squares) and each reference node (blue point). In the third step \ding{194}, we assign a weight to each point in the grid that describes the probability of reaching the point from the reference in the measured time, darker colors represent a higher weight. In the final step \ding{195}, we sum the weights of the grid points within each country, and pick the country with the largest sum (Italy, in this example).}
    \label{fig:mass_process}
\end{figure*}

In localization problems, the specific geographic coordinates are often part of an area that has a semantic significance, \eg, a country, jurisdiction, or zone under the control of a given actor. In this case, all the points within a zone are considered equivalent. In addition to the coordinates that best approximate the node's location, \verloc can ascertain whether the node is located within the zone $z_i$ that contains the claimed location $loc_i$. The process illustrated in Figure~\ref{fig:mass_process} consists of the following steps.

\noindent \textbf{Derive Target Area.}
In a first step, \verloc translates the measured RTTs into a vector of distances. Transmission speeds on the Internet range from~\SI{0.22}{c}~to~\SI{0.67}{c}~\cite{Li2017DeTorPA,katz2006towards}. In order to find the \emph{largest} target area that could possibly meet the constraints of all measurements, we use an upper bound transmission speed of $\nicefrac{2}{3}\cdot c$ to compute the distance $dist_{max}(n_i \leftrightarrow r_j) = \nicefrac{2}{3} \cdot c \cdot RTT(n_i \leftrightarrow r_j)$. 

We sort the distances in ascending order and pick the first element in this list, \ie, the reference appearing to be closest to the target node $n_i$. We ``draw a circle'' around this first reference of radius $dist_{max}(n_i \leftrightarrow r_j)$. The circle describes the area that could have been reached in the measured time. This circle is an initial area where $n_i$ must be located, which is further narrowed down with each additional reference. 

\noindent \textbf{Shrink Target Area.}
We then proceed iteratively with the next references of the sorted list. For each reference $r_j$, we draw a new circle around $r_j$ of radius $dist_{max}(n_i \leftrightarrow r_j)$ and compute the \emph{intersection} with the previous circles. In this iterative process, we narrow down the target area step by step and exit the process when new references do not shrink the target area any further. The approach is robust to network distortions, as occurrences of high background noise lead to longer distances $dist_{max}(n_i \leftrightarrow r_j)$, whose resulting intersections are not overly restrictive. 

\begin{figure}[t]
	\centering
	\includegraphics[width=\columnwidth]{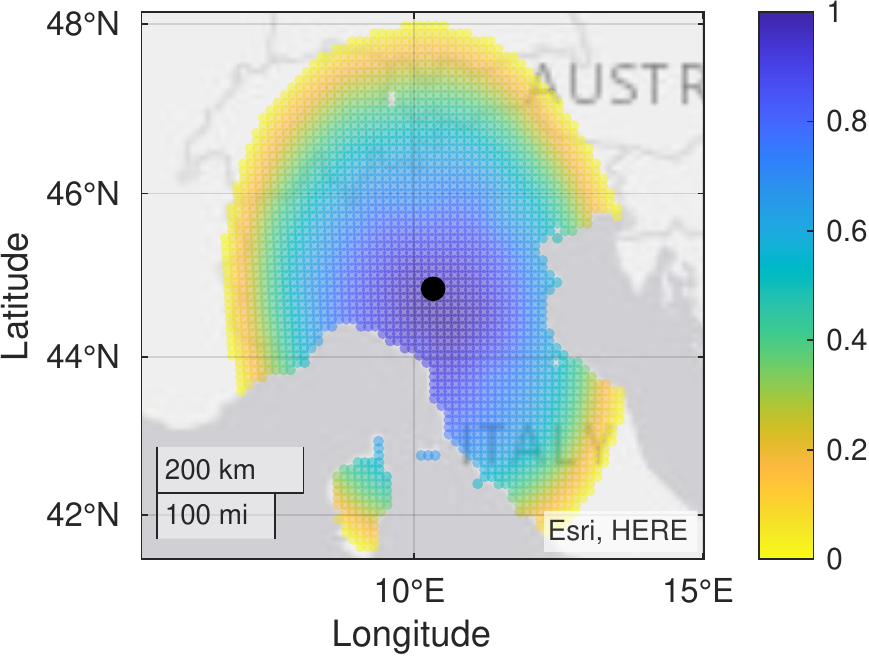}
	\caption{Sample Target Grid. The black point is the location of $n_i$ (ground truth); the colored area shows the weighted grid; darker colors indicate a higher likelihood score for a point in the grid. For the verification decision, we pick the country with the largest sum of likelihood scores.}
    \label{fig:pb_dist_mass}
\end{figure}



\noindent \textbf{Apply Grid.}
In the final step, we apply a grid to the target area resulting from all intersections. We compute a likelihood score for each point in the grid based on $\Delta_j$, which expresses how consistent that point's location is with the measured RTTs. We normalize the scores to obtain a probability distribution, and then sum the scores of the points within each zone that overlaps with the target area. The zone that accumulates the highest mass is then compared to the claimed zone for the verification decision. In the example shown in Figure~\ref{fig:pb_dist_mass} \verloc would output a positive zone verification if the node has claimed to be in Italy, and negative otherwise. Note that it is also possible for \verloc to output the probability score of each zone instead of just the zone with the maximum score.



\subsubsection{Confidence Scores}\label{subsubsec:confidence_scores}

Confidence scores express the degree to which measured RTTs are consistent with the claimed locations of all nodes given the propagation model. This allows to reject decisions where the evidence is inconclusive regarding the node's location. 
In the following, we introduce the \emph{generic} concept of confidence scores, whose parameters we later adjust to detect adversarial timing manipulations (\S\ref{sec:attacks}).

\noindent \textbf{Speed Bounds.} We consider bounds $b_l$ and $b_u$ that define the minimum and maximum propagation speeds for a transmission. The upper bound $b_u = \nicefrac{2}{3}\cdot c$ serves as a sanity check and describes the maximum transmission speed that we usually observe on the Internet. The lower bound $b_l$ is the lower \SI{95}{\percent} confidence bound of the speed fit (\S\ref{subsec:prelim:propagation}). A tolerance factor $ \tau $ accounts for transmission noise that slows down packets beyond the considered speed limit:

\begin{equation}
    b_l = l(x) (1 - \tau) \hspace{0.8cm} 0\leq \tau\leq 1
\end{equation}

\noindent \textbf{Apply Confidence Scores.} 
Consider a target node $n_i$ with a set of references $r_j \in R_i$ and reported timings $RTT(n_i\rightarrow r_j)$ and $RTT(n_i\leftarrow r_j)$ for each direction. The node's confidence score $c_i$ represents the percentage of reference measurement pairs within bounds, with the highest possible score being $1$ and the lowest $0$. More precisely, we count a $1$ for every pair of measurements that stays within the bounds and a $0$ for every pair where at least one direction ($RTT(n_i\rightarrow r_j)$ or $RTT(n_i\leftarrow r_j)$) violates at least one bound ($b_l$ or $b_u$). We then normalize the count dividing by the number of references $|R_i|$. 

A threshold can then be used to either accept or reject $n_i$'s localization results based on its confidence score $c_i$. In Section~\ref{sec:attacks} we show how confidence scores can be used as countermeasure to distinguish benign from manipulated measurements. 

\section{Performance Baseline}\label{sec:simulation}
We begin our evaluation of \verloc with a study of its performance baseline in a non-adversarial simulation setup.

\subsection{Experimental Setup}\label{subsubsec:simulation_network_setup}

We evaluate \verloc with simulations that account for the propagation characteristics of real-world networks. The procedure includes a preparation phase, where we randomly generate a network, and a simulation phase, where we apply \verloc.

\noindent \textbf{Preparation.} We generate a network of $N$ nodes $n_i$, each of which has a randomly chosen true location $loc_i$ in a zone $z_i$ among fifteen of the most populated European countries: Germany, France, United Kingdom, Italy, Spain, Ukraine, Poland, Romania, Netherlands, Belgium, Czech Republic, Hungary, Austria, Switzerland, and Slovakia. Based on these randomly generated locations we generate a propagation matrix that contains pairwise timing measurements for all possible pairs of nodes in the network. We sample the transmission speeds, times, and noise from our empirical propagation model (\S\ref{subsec:prelim:propagation}).

\noindent \textbf{Simulation.} We first generate the set of references $r_j \in R_i$ for each node $n_i$. For this, we pick uniformly at random $t$ references out of the available $N-1$ nodes and extend the set to ensure that all measurements are symmetric (\cf~Alg.~\ref{alg:extension_references}). For each measurement pair we look up timings from the pre-computed propagation matrix. We then apply the method described in Section~\ref{subsubsec:system_estimate_locations} to estimate $n_i$'s location $\widehat{loc_i}$ and the method of Section~\ref{subsubsec:location_verification} to verify its zone, as defined by country borders. 

\subsection{Metrics}\label{subsubsec:simulation_metrics}

We use two metrics to evaluate the performance baseline. First, we measure the \emph{location error} of a node $n_i$ as the great circle distance between the estimated and actual node locations, $dist(loc_i, \widehat{loc_i})$. Second, we compute the \emph{zone verification rate} as the fraction of nodes for which the highest weight zone matches the ground truth of the node's location zone.

\subsection{Experiments}\label{subsec:experiments_pb}

\noindent \textbf{Number of References.} 
\label{subsubsec:pb_num_refs}
As initial step to adjust \verloc's parameters, we analyze how the number of references influences node localization accuracy and zone verification rates. The number of references $|R_i|$ of node $n_i$ strikes a tradeoff between the overhead and performance of \verloc. References must be picked randomly to prevent attacks. Thus, we cannot optimize the choice of $R_i$ to maximize proximity to $n_i$ or diversity of directions, which would increase localization accuracy. Increasing the number of references is the next best option to ensure diversity of measurement directions and to level out noise.



\begin{figure}[t]
	\centering
	\includegraphics[width=.9\columnwidth]{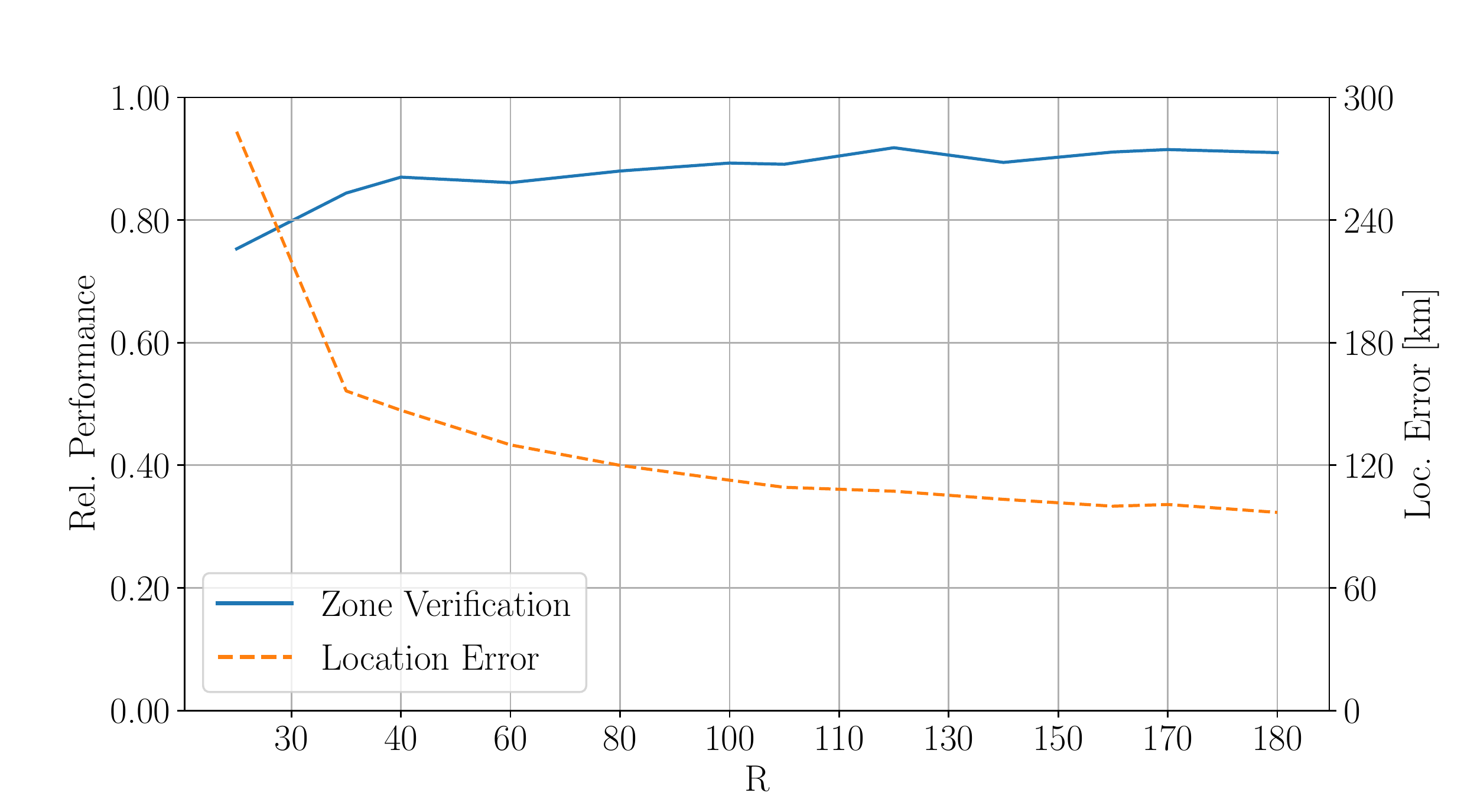}
	\caption{Performance for Localization and Verification. The zone verification rate (left y axis) shows the fraction of correct zone verifications. The localization error (right y axis) shows the distance between the estimated and true node locations.} 
    \label{fig:pb_mass_distance}
\end{figure}

To find a suitable target number $R$ of references, we compute the average location error and zone verification rate for different values of $R$, and show the results in Figure~\ref{fig:pb_mass_distance}. We observe a significant performance improvement in the range of $R=10$ to $R=80$ references, which then offers diminishing returns for larger values of $R$. For our simulations we choose $R=80$ as lower bound for the number of references, and set parameter $t=50$ so that reference sets are larger than $R=80$ in at least \SI{98}{\percent} of cases, \ie, $|R_i| \geq R$ for most nodes (\cf~Eq.~\ref{eq-R-t}). In the real-world experiments presented later (\S\ref{sec:prototype}) we test both $R=40$ and $R=80$ and find that the difference in localization accuracy is less than $0.5$ km. 
 


\noindent \\\textbf{Baseline Parameter Values.}
\label{subsec:parameter_setup}
We document the \verloc parameters in Table~\ref{tbl:parameter_setup}. The first two blocks describe the network setup and the number of references we determined in the performance baseline. The following two blocks are dedicated to the performance of \verloc in an adversarial setting, which we discuss in the following section.

\begin{table}[t]
	\centering
	\caption{Simulation Parameter Setup.}
	\begin{tabular}{llll}
        \toprule
		\textbf{Parameter} & \textbf{Notation} & \textbf{Value} &\textbf{Section}\\
		\midrule
		Nr Network Nodes & $N$ & $1000$ & \multirow{7}{*}{\S\ref{subsubsec:simulation_network_setup}}\\
		Network Node & $n_i$ & -- & \\
		Keys & $(pk_i,sk_i)$ & -- &  \\
		True Physical Loc. & $loc_i$ & -- &  \\
		Estimated Loc. & $\widehat{loc_i}$ & -- &  \\
		Claimed Loc. & $loc^A_i$ & -- &  \\
		True Physical Zone & $z_i$ & -- &  \\
		Estimated Zone & $\widehat{z_i}$ & -- &  \\

		\midrule

		Base References & $t$ & $50$ & \multirow{2}{*}{\S\ref{subsubsec:pb_num_refs}}\\
		Reference Set & $r_j \in R_i$ & $|R_i| \geq 80$ \\

		\midrule

		Nr Adversarial Nodes & $|A|$ & $50$ .. $300$ & \multirow{2}{*}{\S\ref{subsec:atk_exp_setup}}\\
		Adversarial Nodes & $a_j \in A$ & -- & \\
		Claimed Nodes & $c_k \in C$ & $C \subseteq A$ & \\

		\midrule

		Confidence Threshold & $\upsilon$ & \num{0.2} & \multirow{2}{*}{\S\ref{subsubsec:atk_conf_optimization}}\\ 
		Tolerance Factor & $\tau$ & \num{0.01} & \\
		
		\bottomrule
	\end{tabular}
\label{tbl:parameter_setup}
\end{table}

\begin{figure}[t]
	\centering
	\includegraphics[width=.9\columnwidth]{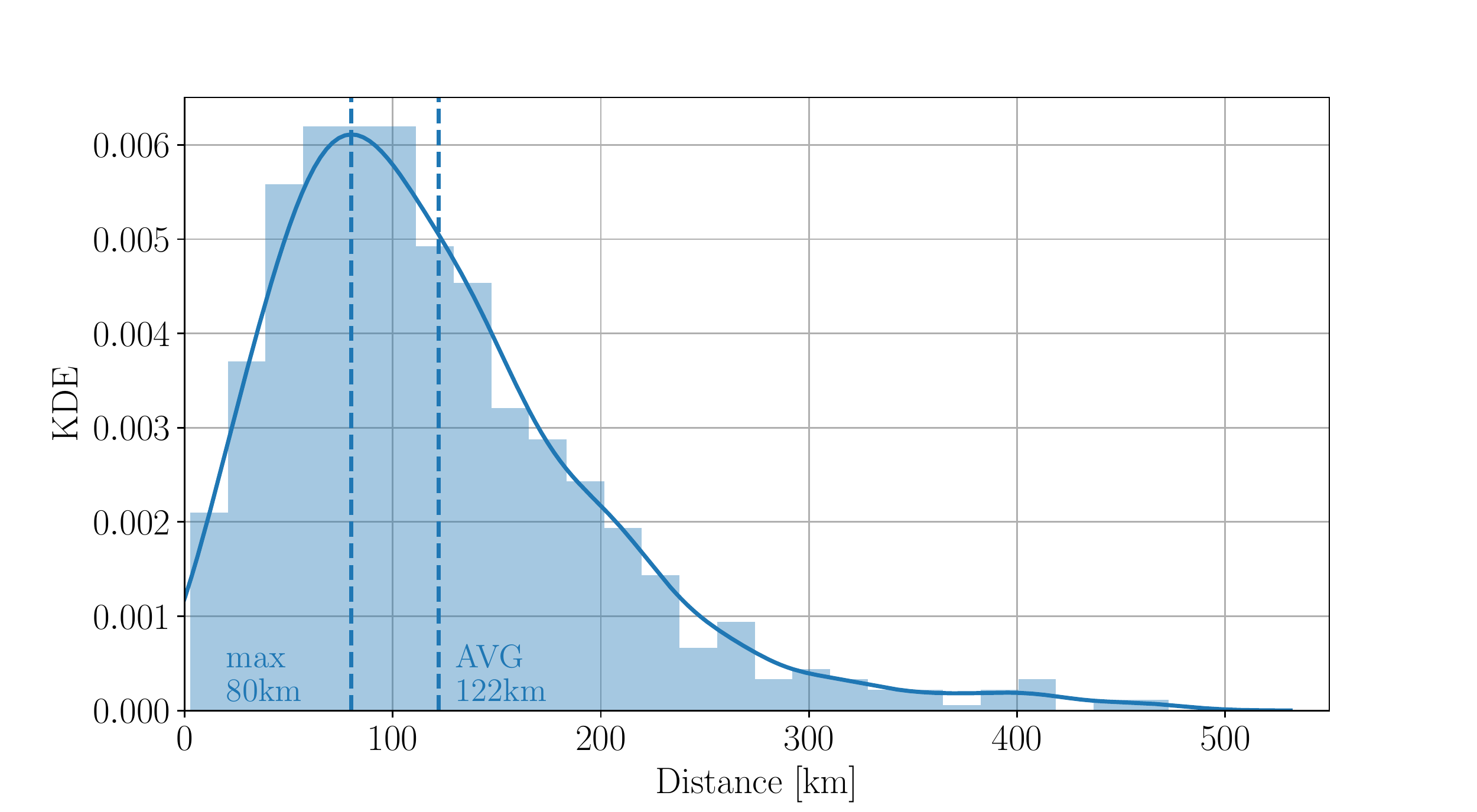}
	\caption{KDE Localization Error: Distribution of distances between the estimated and the true location of nodes.}
    \label{fig:geo_histfit-sim}
\end{figure}

\noindent \textbf{Localization and Zone Verification Results.} 
In our experiments the median localization error is \SI{103}{\km} and the average is \SI{122}{\km}. The distance between true and estimated locations is distributed as shown in Figure~\ref{fig:geo_histfit-sim}. To get a sense of how these distances compare to European country sizes, we empirically evaluate the likelihood that the error will move the node across a country border. As expected, shifts between small neighboring countries are more likely to happen, but \verloc still achieves on average \SI{92}{\percent} accuracy when verifying the country in Europe where a node is located. 
We note that experiments in the wild outperform simulation results (\S\ref{subsec:results-real-world}), confirming that our simulations represent a particularly difficult deployment scenario. 

\section{Security Analysis}\label{sec:attacks}

We consider an adversary that participates in the network with several malicious nodes. The adversary claims false locations (different from the nodes' true physical locations) for a subset of those nodes and manipulates measurements consistently with the fake claimed locations. Claiming false locations undermines the localization and verification capabilities of \verloc compared to the baseline. We extend here the initial network model (\S\ref{subsec:problem_statement}) to account for adversarial setups. 

\subsection{Adversarial Timing Manipulations}\label{subsec:atk_timing_manipulations}

We consider a network of $N$ nodes, of which a subset $A$ of nodes is adversarial, $A\subset N$. The adversary claims false locations $loc^A_k \neq loc_k$ for a subset $C$ of adversarial nodes, $C \subseteq A$. The adversary \emph{controls} all nodes in $A$ and can thus manipulate timings whenever an adversarial node participates in a measurement. 

Given a malicious target node $c_k \in C$ and its reference set $R_k$, there are three possible scenarios for the target and reference pairings $c_k \leftrightarrow r_j$.

\noindent \textbf{Perfect Manipulation.}
A \emph{perfect manipulation} is possible when both the target node $c_k$ \emph{and} the reference $r_j \in R_k$ are under adversarial control, \ie, $c_k \in C$ \emph{and} $r_j \in A$. In this case the adversary contributes spoofed times $RTT(c_k \rightarrow r_j)$ and $RTT(c_k \leftarrow r_j)$ for both directions. The spoofed timings match the expected propagation time for the claimed locations:

\begin{equation}
    RTT(c_k \rightarrow r_j) = RTT(c_k \leftarrow r_j) = \frac{dist(loc^A_k, loc^A_j)}{f(dist(loc^A_k, loc^A_j))}
\end{equation}

More precisely, the adversary first computes the distance between the \emph{claimed} location $loc^A_k$ and the adversarial reference node $loc^A_j$. Note that $loc^A_j = loc_j$ if $r_j \in A$ but $r_j \not\in C$, while $loc^A_j \neq loc_j$ if $r_j \in C$. 
This distance serves as an input to the empirical speed function $f(x)$ (\cf~\S~\ref{subsec:prelim:propagation}). We assume that all the parameters of \verloc are known to the adversary, who can apply the propagation model to compute timings that match the claimed distance. The adversary can slightly alter reported RTT values with noise to avoid submitting suspiciously identical numbers. 


The following two scenarios cover cases in which the reference node is \emph{not} adversarial, \ie, $c_k \in C$ but $r_j \not\in A$. 

\noindent \textbf{Slowing Down.}
If the claimed location $loc^A_k$ of $c_k$ is \emph{further} away from reference $r_j$ than the true location $loc_k$, \ie, if $dist(loc^A_k,loc_j) > dist(loc_k,loc_j)$, then it is possible for the adversary to \emph{slow down} incoming timing probes $(c_k \leftarrow r_j)$. Slowing down means that the adversary delays its response to the incoming probe in order to bring the $RTT(c_k \leftarrow r_j)$ measured by $r_j$ close to the value that would be observed if $c_k$ was indeed placed in $loc^A_k$. 

As in the previous case, we consider the adversary is able to tamper with timings \emph{in both directions} to perfectly fit the claimed location. In one direction, the adversary contributes a spoofed timing, while in the other direction it adds latency to manipulate the timing reported by the (honest) reference node $r_j$.

\noindent \textbf{No Manipulation.}
If the claimed location $loc^A_k$ of $c_k$ is \emph{closer} to the benign reference $r_j$ than its true location $loc_k$, \ie, if $dist(loc^A_k,loc_j) < dist(loc_k,loc_j)$, then the adversary \emph{cannot} manipulate the measurements taken and reported by $r_j$, as he cannot speed up probes~\cite{kohls19multi} or reply before receiving the reference's probe and seeing the included nonce.

\subsection{Experimental Setup}\label{subsec:atk_exp_setup}
In an adversarial setup, we are interested in the general system performance (accuracy of localization and zone verification rate) for honest nodes, as well as the number of successfully claimed false locations. To this end, we adjust the \emph{network setup} and applied \emph{metrics}.

\noindent \textbf{Network Setup.} 
As in the performance baseline experiments, we simulate networks of $N=1000$ nodes placed in random locations across Europe and use a reference set size of $R=80$.

We additionally define a subset $A$ of adversarial nodes that can manipulate measurements and a subset $C \subseteq A$ of nodes for which the adversary claims false locations. We randomly pick fake claimed locations but ensure that the falsely claimed location is in a different country than the actual location of the adversarial node. 

We first run a simulation considering true locations for all nodes. We then substitute the propagation times for all measurements in which the adversary can either apply a perfect manipulation or slow down incoming timing probes. We re-apply the localization and verification methods for all the nodes that were affected by the adversarial activities. This includes all nodes $c_k \in C$ with false claimed locations, but also all the benign nodes in their reference sets $R_k$. We do so to examine all the discrepancies that the adversary introduces when introducing bogus locations and measurements

\noindent \textbf{Metrics.} 
To measure the performance of \verloc in an adversarial setup, we extend the initial performance metrics (\S\ref{subsubsec:simulation_metrics}) with a set of \texttt{true/false}, \texttt{positive/negative} results. In contrast to the benign setup, these results now include accept/reject decisions for individual nodes.

We denote a \texttt{positive} decision as an accept, \ie, a decision where the confidence score is sufficiently high; a \texttt{negative} is a rejected decision where the confidence falls below a defined threshold $\upsilon$. Furthermore, we treat a result as \texttt{true} whenever it matches the ground truth and as \texttt{false} when it contradicts the ground truth:

\begin{itemize}
    \item \textbf{TP} A true positive decision denotes an \emph{accept} for a node that reports its true location $loc_i \in z_i$, meaning that \verloc believes the node to be in the correct zone $z_i$. 
    \item \textbf{TN} A true negative decision denotes a \emph{reject} for an adversarial node claiming to be at a false location $loc^A_k \in z_k^A$ such that $z_k^A \neq z_k$. 
    \item \textbf{FP} A false positive decision denotes an \emph{accept} for a node with a false claimed location $loc^A_k \in z_k^A$, meaning that \verloc believes the node to be in the falsely claimed zone $z_k^A$. 
    \item \textbf{FN} A false negative decision denotes a \emph{reject} for a node whose claimed location $loc_i \in z_i$ was actually true.
\end{itemize}

In the following, we use a \emph{confidence score} to identify fake node locations. We then analyze the attack performance for an increasing number of adversarial nodes to test the \emph{breaking point} of the system.

\subsection{Attack Success}
In a first step, we analyze the success of the adversary in claiming false locations for some of its nodes, considering a network of $N=1000$, where each has at least $R=80$ randomly chosen references. To this end, we allow the adversary to control $|A|=50$ randomly chosen nodes and evaluate the attack success for an increasing number of false location claims $|C|=5..50$. Note that in the case of $|C|=50$ the adversary is claiming false locations for all adversarial nodes, \ie, $C=A$. 

\begin{figure}[t]
	\centering
	\includegraphics[width=.9\columnwidth]{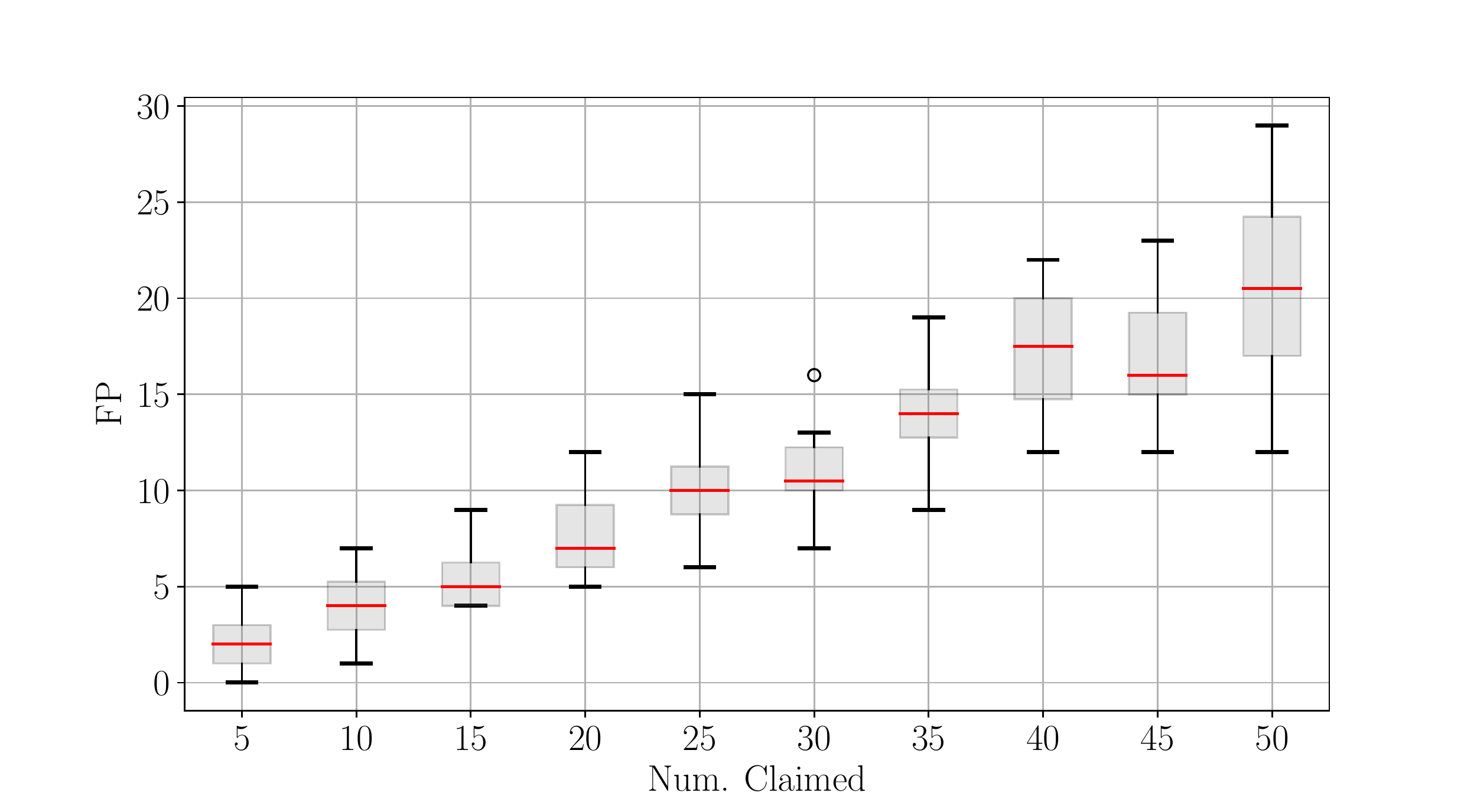}
	\caption{Adversarial success for a given number of false claimed locations (\num{20000} samples).}
    \label{fig:unprotected_system}
\end{figure}

We show the number of successfully verified false locations in Figure~\ref{fig:unprotected_system}, where we can see that in the absence of countermeasures the adversary is successful between a third and half of the times. 

\subsection{Confidence Scores}\label{subsubsec:atk_conf_optimization}

To harden \verloc and mitigate the attack success, we use the confidence scores introduced previously (\S\ref{subsubsec:confidence_scores}). We define a \emph{reject threshold} $\upsilon$ to distinguish between benign and malicious location claims. We further define a \emph{tolerance factor} $\tau$ to account for background noise that disturbs the end-to-end timings. These are the two main parameters that determine accept and reject decisions based on the confidence scores.

The tolerance factor $\tau$ defines how much noise is tolerated, \ie, it relaxes the lower bound $b_l$ on the speed, accepting even slower transmissions. Too many transmissions being too slow is precisely a distinguishing characteristic of false claimed locations. Whenever a false claimed location is \emph{closer} to a reference $r_j$ than the true location, the adversary is unable to manipulate the measured timing to make it shorter. Consequently, the RTT reported by the reference $r_j$ will appear as a very slow (noisy) transmission. 

The threshold $\upsilon$ defines the minimum decision confidence score required by \verloc to accept a claimed location as verified. This parameter influences the tradeoff between false positive and false negative rates. An overly restrictive $\upsilon$ will reject many decisions, including those of benign nodes for which measurements are simply noisy. On the other hand, a very lax $\upsilon$ increases false positives, correctly verifying a larger number of honest node locations at the cost of also accepting some false locations as correct.

\noindent \textbf{Tolerance Factor ($\tau$).}
To evaluate how the tolerance factor $\tau$ influences the overall performance of \verloc, we test values in the range of $\tau=0.005..0.025$ considering a decision threshold $\upsilon=0.2$.
In our evaluation, we first study which configurations prevent the adversary from claiming \emph{any} false location, \ie, cases in which $FP=0$. 
We find that there is minimal variation for the accept and TP rates. More precisely, the acceptance rates are in the range of \SIrange{95}{96}{\percent}, and the TP rates achieve \SIrange{86}{87}{\percent} within the accepted decisions. For the simulation experiments we use a tolerance factor of $\tau=0.01$.

\begin{figure}[t]
	\centering
	\includegraphics[width=.9\columnwidth]{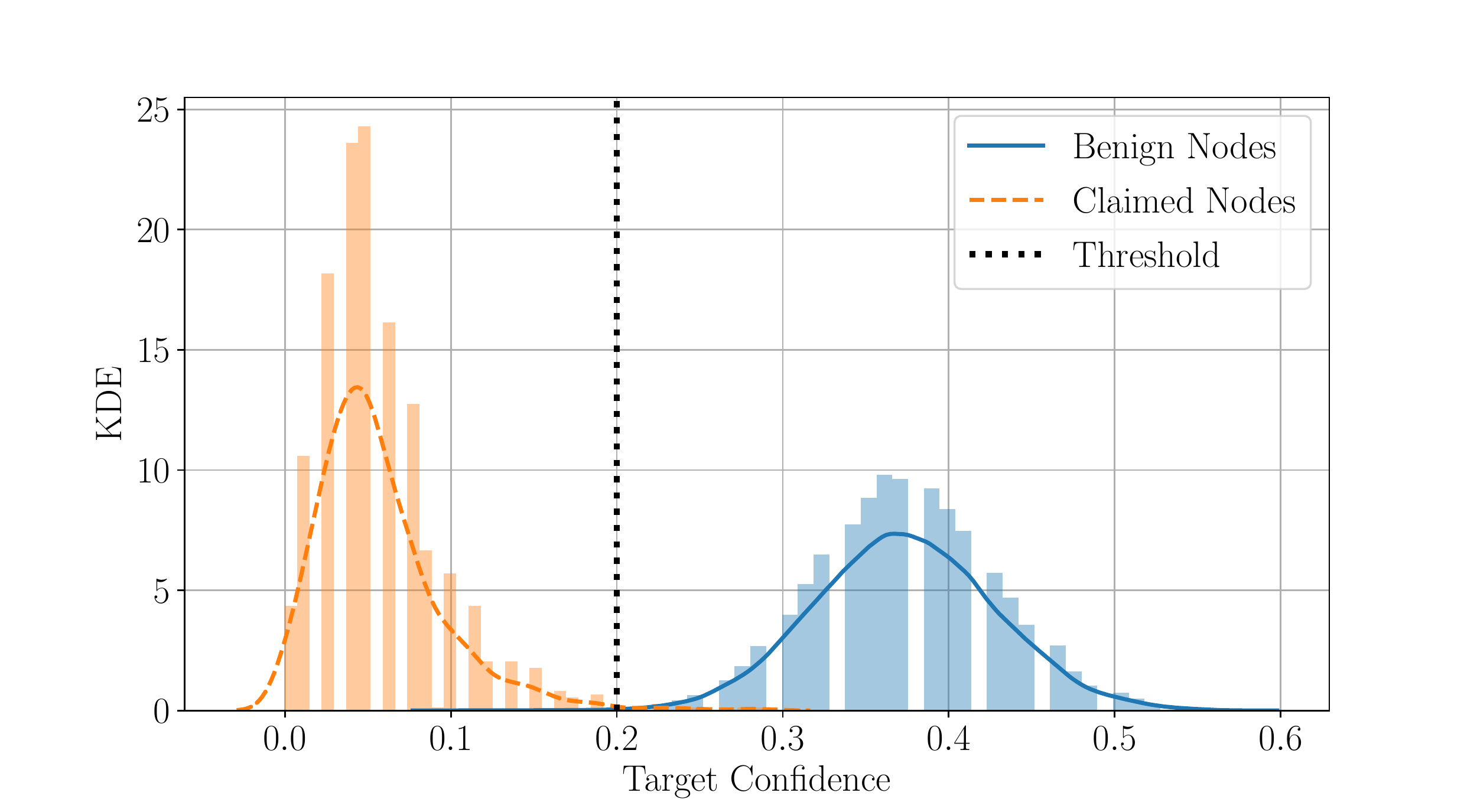}
	\caption{Distribution (kernel density estimate) of confidence scores. Benign (blue) and malicious (orange) nodes in a setup with $|C|=|A|=50$ adversarial nodes claiming false locations. We choose a threshold $\upsilon=0.2$ to distinguish both groups.}
    \label{fig:confidence_score_curves}
\end{figure}

\noindent \textbf{Decision Threshold ($\upsilon$).}
\label{subsubsec:confidence_decision_threshold}
The decision threshold $\upsilon$ defines the minimum required confidence to accept a decision.
Figure~\ref{fig:confidence_score_curves} shows the distribution of confidence scores for adversarial and benign nodes considering a scenario where the adversary controls $50$ nodes, all of which claim false locations. As we can see in the figure, both groups can be perfectly distinguished with a decision threshold of $\upsilon=0.2$, as there is no overlap between both distributions. We thus choose this value for $\upsilon$.

\subsection{Breaking Point of \verloc}

We have shown that \verloc can reliably distinguish between manipulated and benign node decisions. This prevents adversarial success while maintaining a reliable verification and localization performance in situations where the adversary controls a limited percentage of nodes. However, we are also interested in identifying the \emph{breaking point} of the system, \ie, the required amount of adversarial resources that degrades the performance of \verloc to unacceptable levels. To this end, we gradually increase the fraction of adversarial control in the network (\cf~Table~\ref{tbl:atk_parameter_selection}). We observe: 

\begin{figure}[t]
	\centering
	\includegraphics[width=.9\columnwidth]{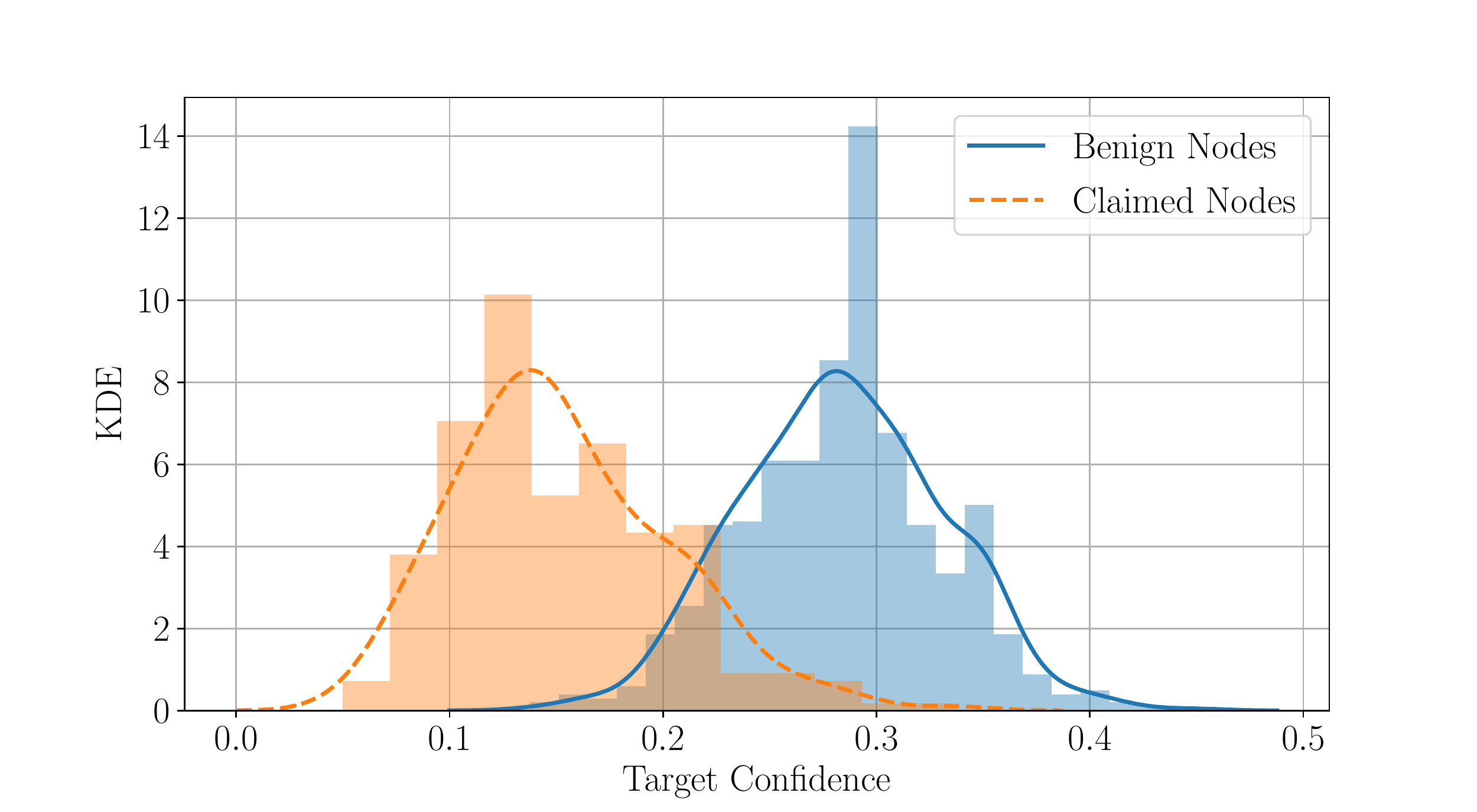}
	\caption{Distribution of confidence scores for a setup with $250$ out of $1000$ ($25\%$) malicious nodes, all claiming false locations.}
    \label{fig:break_confidence}
\end{figure}

\begin{itemize}
    \item \textbf{Confidence Score Distribution.} An increasing number of adversarial nodes with false claimed locations introduces distortions in many measurements, and the manipulations lead to network-wide inconsistencies between claimed locations and reported timing measurements. Figure~\ref{fig:break_confidence} shows how the confidence scores begin to overlap for benign and malicious nodes at \SI{25}{\percent} compromise.

    \item \textbf{False Positives.} \verloc begins to output FP results when reaching around a third of adversarial nodes with false location claims. While the adversarial success is limited in this setup ($FP=3$), it indicates that at this point the adversary is capable of compromising the verification process to successfully claim false locations. Note that a FP only occurs when (1) the estimated zone $\hat{z_i}$ coincides with the zone that contains the location $loc^A_i$ claimed by the adversary, and (2) the adversarial node has a confidence score larger than $\upsilon$. A confidence score above $\upsilon$ does \textit{not} lead to a FP if the estimated and claimed zones do not coincide. 
    
    \item \textbf{False Negatives.} False negatives do not compromise the system's security directly but they have the risk of unfairly rejecting honest nodes from the system. Furthermore, an increasing FN rate indicates that \verloc looses the ability to make reliable accept and reject decisions. We see that FN begins to increase when adversaries control \SI{20}{\percent} or more of the network. 
\end{itemize}

\begin{table}[t]
	\centering
	\caption{System Breaking Point.}
    \begin{tabular}{cccccc}
        \toprule
         \textbf{Claimed} & \textbf{Reject} & \textbf{TP}  & \textbf{FP} &  \textbf{FN} & \textbf{Recall}\\
        \midrule
        \SI{5}{\percent} & 50 	& 815	& 0	& 0	    & 1.00	    \\
        \SI{10}{\percent} & 102 	& 756	& 0	& 2	    & 0.99	\\
        \SI{15}{\percent} & 152 	& 691	& 0	& 3	    & 0.99	\\
        \SI{20}{\percent} & 214 	& 627	& 0	& 23	& 0.96	\\
        \SI{25}{\percent} & 284 	& 555	& 0	& 58	& 0.91	\\
        \midrule
        \SI{30}{\percent} & 335 	& 476	& 3	& 105	& 0.82	\\
        \SI{35}{\percent} & 352 	& 401	& 3	& 151	& 0.73	\\
        \bottomrule         
        \end{tabular}
	\label{tbl:atk_parameter_selection}
\end{table}

\noindent \textbf{Conclusion.}
The confidence scores of \verloc allow distinguishing true and false location claims. This protection mechanism is robust to adversaries that control up to about \SI{20}{\percent} of nodes in the network while still providing high performance rates for all remaining honest nodes. Note that these results correspond to a network setup restricted to Europe, which is a challenging use case (\S\ref{subsec:discussion_attack_variants}).

\subsection{Additional Threats}
\label{subsec:discussion_attack_variants}

\subsubsection{Framing Benign Nodes}

Instead of trying to successfully claim false locations (FP) for malicious nodes, the adversary can attempt to frame benign nodes as claiming a false location (FN) by, \eg, contributing timings that violate bounds and reduce the confidence score of an honest node. An adversary that controls a fraction $\alpha=\frac{|A|}{N}$ of all the nodes is on average able to manipulate a fraction $\alpha$ of an honest node's reference measurements, in one of the two directions.\footnote{The exact number of corrupt references in a node's set is given by a hypergeometric distribution with population size $N$, $R$ draws, and $|A|$ special (malicious) objects. The variance of such distribution is low for a large $R$, making it unlikely that the adversary controls much more than a fraction $\alpha$ of references in a randomly chosen subset.} This is in contrast with claiming false locations, where the adversary controls \textit{all} the reference measurements in at least one direction, and a fraction $\alpha$ in \textit{both} directions. 

An adversary that controls a fraction $\alpha$ of a node's references can \textit{at most} lower the confidence score of the node by $\alpha$, by making $\alpha \cdot R$ measurements that would otherwise be within bounds to be out of bounds.  Considering the results shown in Figure~\ref{fig:confidence_score_curves} for benign nodes, more than half the references of a node fail the bounds test in the absence of attack, meaning that the adversary in practice will be only able to lower an honest node's score by less than $\alpha$. Considering a threshold $\upsilon=0.2$ for the confidence scores, an adversary has to control at least a fraction $\alpha=0.2$ to start bringing down below $\upsilon$ the confidence score of a non-negligible fraction of benign nodes. 

Note that, because measurements are symmetric, a bounds violation in a pairwise measurement affects both nodes involved, and thus the scores of the adversarial nodes themselves become lower as they attack more targets, meaning that the adversary has to trade scalability with detectability of the attack. 

\subsubsection{Strategic Locations for Adversarial Nodes}

\NEW{5}{An adversary can improve the attack success by strategically positioning its nodes. For this strategic placement, two key characteristics can be exploited. First, zone verification errors where a node is believed to be in a different zone are more likely to occur in \textit{small zones}, such that an error of less than a hundred kilometres is enough to shift the node to a different zone. 
Thus, an adversary placing the nodes in small zones can more plausibly claim that the observed discrepancies are due to natural network effects and noise rather than malicious manipulation. Within the localization constraints the adversary is subject to, placing the malicious nodes as close as possible to the border of the claimed zone increases the chances of adversarial success. Note that regardless of zone boundaries, it is always easier for an adversary to plausibly claim being at a nearby location rather than a faraway location. } 

\NEW{5}{Second, we investigate how the \textit{directional diversity} of a reference set influences the accuracy of localization. In the best case, a node's references are situated in all directions to provide the highest possible measurement diversity. To illustrate why directional diversity is important, consider a case where multiple measurements have a lot of noise (extra latency). If the noisy measurements come from opposite directions, the discrepancies are leveled out and \verloc arrives at a good estimation. In contrast, noisy measurements from a single direction push the estimated target node further away from its actual location.
In our experiments, \SI{95}{\percent} of failed verification for nodes in Spain, which is located in the SW corner of Europe, are associated with a unidirectional distribution of references, whereas this issue appears related to just \SI{20}{\percent} of nodes in Romania, which has a more central location in the continent. Therefore, an adversary can more plausibly claim that a failed verification is due to natural causes when claiming locations with less directional diversity. Note also that the adversary has more influence on \verloc's results for a node when it is the only reference providing measurements for that node from a certain direction.}

\NEW{5}{The study of adversarial node positioning strategies to achieve concrete objectives under specific constraints is left for future work. Such further research may, for example, evaluate strategies for successfully convincing the network that a number of adversarial nodes are located within a specific country (that, e.g., has a strong rule of law and favorable legislation, which increases trust in those adversarial nodes), when they are actually located somewhere else that is at a certain distance; or strategies for downgrading the confidence score of specific targets among the honest nodes. }


\subsubsection{Adversaries in the underlying network}

\NEW{5}{Network adversaries placed in between two nodes may intercept probes and respond to them, causing nodes to measure RTTs that are impossibly small given the actual distance between them. Security can be strengthened towards such adversaries if the two nodes, who can authenticate each other using the public keys in the node descriptors, establish a shared secret $s$ before exchanging probes, \eg, with an authenticated Diffie-Hellmann key exchange. Instead of simply copying the nonce in their response, nodes respond to probes with a hash of $s$ concatenated with the nonce. A man-in-the-middle adversary who is physically between the two nodes cannot fake a shorter distance without access to $s$. Note that the time required to compute this hash adds to the measured RTT and thus needs to be factored in when building the propagation model. If the hash computation time is highly variable from one node to another, this adds noise that may decrease localization accuracy; if on the other hand the hash computation time is rather constant across servers, it can be easily accounted for in the propagation model without an impact on localization accuracy.}

\NEW{5}{Alternatively, a network adversary can always slow down probes it intercepts. Note that this gives no additional advantage to an adversary that controls one of the nodes involved in the probe, who already has the ability to slow down the probe at will; but it does enable the adversary to delay probes sent between honest nodes. The effect of such an attack is to lower the confidence score of honest nodes, who now appear to be far away from many other nodes, or even all other nodes if the adversary fully controls the network connection of the victim. Note that distinguishing such delays from natural transmission delays due to poor network conditions is non-trivial, as both effects cause a lower confidence score. We argue that this is an acceptable effect, as the confidence scores not only represent adversarial manipulations, but also take into account bad network conditions. In both cases the score represents a lower confidence in the localization result.}

\subsubsection{Compromised Broadcast Channel}

\NEW{5}{\verloc relies on two types of node information to produce results: node descriptors (containing IP addresses, claimed locations and public keys) and recorded measurements. We assume that node descriptors are always securely broadcast even in the absence of \verloc, as otherwise it is easy to completely disrupt the network and any functionality it could offer, beyond \verloc's node localization features. Note that in Tor this information is included in a consensus document signed by all directory authorities, while in Nym it is collectively signed by validators and published in the blockchain. The recorded measurements are however specific to \verloc and not broadcast already as part of basic network orchestration. Therefore, we can expect \verloc implementations such as the one described in the next section, where recorded measurements are made publicly available in ways that are more susceptible to adversarial attacks. An adversary could, for example, show different measurement results to different participants by serving a different result files depending on the IP address of the requester. In this case participants may arrive to different conclusions regarding the localization of some nodes. We note that public web pages with node information such as those maintained by Nodes Guru\footnote{\url{https://nodes.guru/nym/nymworld}} make such attacks detectable, as participants may realize that their locally obtained \verloc results do not coincide with results shared by others in public places. Thus, while such attacks are difficult to prevent in the absence of secure broadcast for measurements, they can be relatively easy to detect (and react to) given active community engagement, discussion and scrutiny of the system.}

\section{Experiments in the Wild}\label{sec:prototype}

As final step we validate \verloc with a real-world experiment where deployed nodes run a simplified prototype implementation.

\subsection{Experimental setup}
\label{subsec:prototype_measurements}

The simplified \verloc prototype implementation was bundled with the Nym network's mixnode code version 0.10.1, released on May 25th. \NEW{3}{Thousands of nodes are actively participating in \verloc measurements and publishing new results every \num{12} hours. More precisely, a list of mixnodes with  IP addresses of all nodes involved in the network is publicly available.\footnote{\url{https://testnet-finney-explorer.nymtech.net/data/mixnodes.json}} Using these IP addresses, we can load the measurement results from a specific port directly at the node \texttt{http://[IP]:8000/verloc}. We have made publicly available the prototype implementation\footnote{\url{https://github.com/katharinakohls/VerLoc}} used to obtain the results shown in this paper. The prototype includes the main implementation of \verloc and a fetch and parsing script to obtain the node measurements. Our results are therefore fully reproducible and the prototype is available to conduct further studies.} 

The implementation functions as follows. As part of Nym's normal features (independently of \verloc), mix nodes periodically download the full list of active nodes in the network, with their public keys and IP addresses. Nodes self-report location as part of their information, typically at the level of nearest town. \NEW{2}{In Appendix~\ref{app:self-reported} we provide details on the quality of self-reported locations.}
A node that runs the \verloc-enabled version of the software sends $200$ ICMP echo requests to each of the other nodes in the network and records the measured RTT values. 
The minimum RTT per reference is made available at a defined port accessible through the node's IP address. We crawl and parse these files to retrieve the measurements. We then locally run our Python scripts on that data to compute localization results. The main differences between the prototype version that is currently deployed and the full proposed \verloc system are: 
\begin{itemize}
    \item The prototype does not publish the measured RTTs in a blockchain but instead makes them publicly available as json files. 
    
    \item In the prototype nodes measure \textit{all} (thousands) other nodes in the network rather than a subset of $40$ to $80$ references. This is to allow us to collect a larger dataset and perform a more thorough evaluation. A complete \verloc implementation conducts fewer measurements (two orders of magnitude less) and thus generates \textit{much} less data so that it is feasible to publish RTT measurements in a blockchain.  
\end{itemize}

\subsection{Results}
\label{subsec:results-real-world}

At the time of submission (June 2021) there are \num{7469} individual nodes in the Nym network. After crawling, parsing, and filtering for nodes running the \verloc-enabled version 0.10.1, we were left with \num{3460} nodes for our evaluation. The main `loss' of nodes running the right software version is caused by problems parsing or mapping the self-reported locations. In the current prototype implementation, node operators provide location as a string that should be the name of a city. We then look up the name in a publicly available list of \num{26569} world cities. We use the coordinates of the center point of the city as an approximation of the node's self-reported location. Unfortunately, the free text leads to parsing errors due to spelling mistakes and formatting issues. Furthermore, if nodes provide the name of a small place not in the list of world cities, our current prototype cannot assign a reported location to the node. In a more advanced version of \verloc, the self-reported location should be standardized, \eg, by providing numerical fields for reporting latitude and longitude instead of a free text field for the location name---or substituted altogether by GeoIP locations. 

For comparison to the simulation results (\S\ref{sec:simulation}) we select nodes in Europe that have a sufficient number of measurements, which results in a subset of \num{943} out of the \num{3460} nodes. The filtered dataset with \num{943} nodes contains \num{852870} individual measurements, with an average of \num{902} references per node. For each selected node, we randomly choose $R$ of its references, discarding and re-sampling if we get a duplicate location already in the reference set. We then retrieve the corresponding measurements from the dataset. 

\begin{figure}[t]
	\centering
	\includegraphics[width=.9\columnwidth]{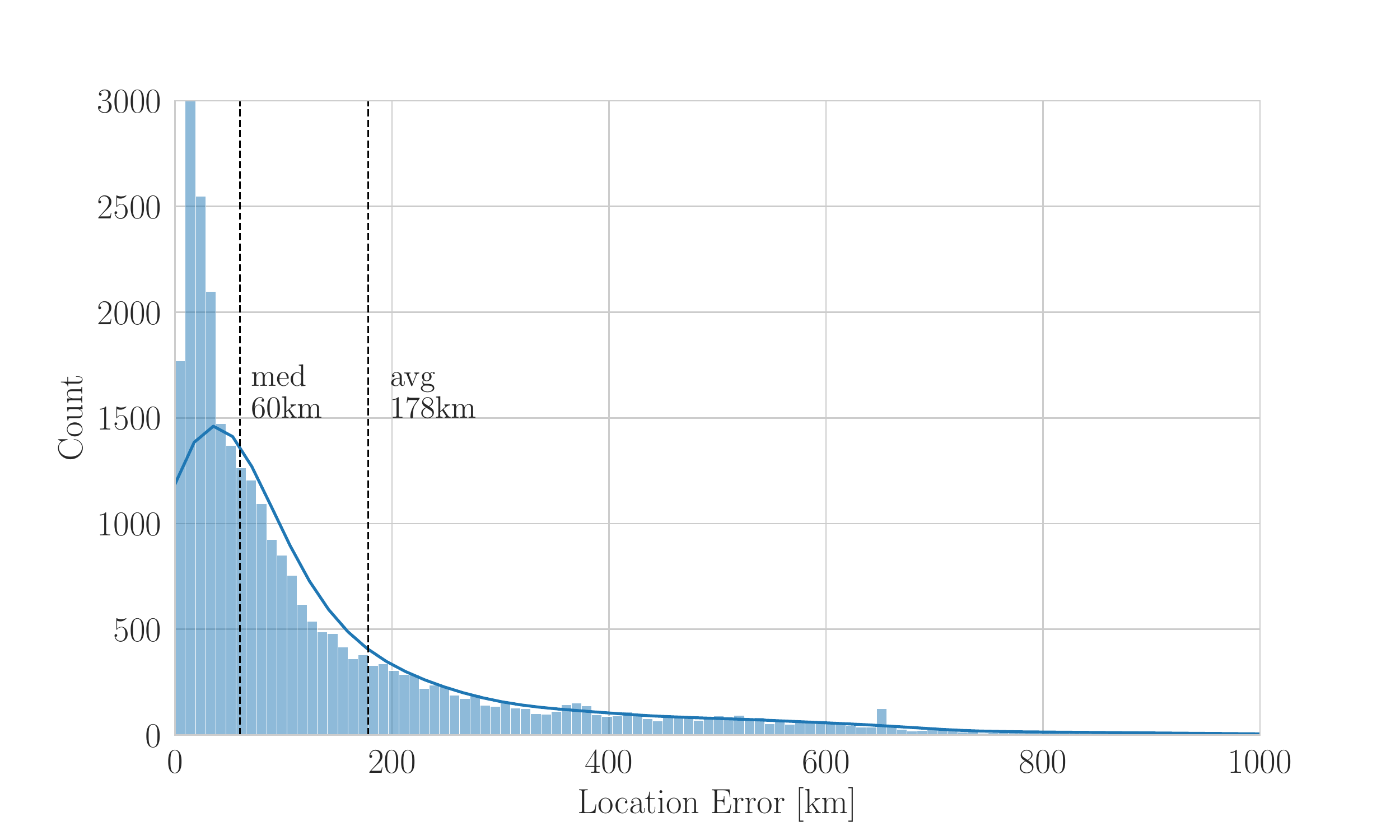}
	\caption{KDE Localization Error: Distribution of distances between the estimated and the true location of nodes.}
    \label{fig:geo_histfit}
\end{figure}

We tested $R=80$ and $R=40$ and the median localization error differed by less than \SI{0.5}{\km}. The results show a median localization error of \SI{60}{\km} and an average of \SI{178}{\km}, which are better results than those obtained in the simulations (\S\ref{sec:simulation}). The distribution of error is shown in Figure~\ref{fig:geo_histfit}. Compared to Figure~\ref{fig:geo_histfit-sim}, it is more skewed, with a larger number of rather accurate localizations and a long tail of error. 
This performance improves when applying confidence scores (\S\ref{subsubsec:atk_conf_optimization}). Although we do not expect an active manipulation of measurements in the collected data set, self-reported locations may be grossly inaccurate in some cases. Confidence scores help filter out nodes with abnormally high rates of bad measurements. When adjusting the tolerance parameter $\tau=0.2$ and keeping the threshold $\upsilon=0.2$, two thirds of the nodes have confidence above the threshold. The median localization error for nodes with a confidence score $c_i>\upsilon$ is \SI{55}{\km}, while the median error for nodes with a confidence score $c_i<\upsilon$ is \SI{256}{\km}.




\section{Overhead of \verloc}

\verloc is intended for networks that already have a system for broadcasting node descriptors, so that this requirement does not impose additional overhead as it piggy-backs on existing infrastructure. The network may also collaboratively generate a periodic random beacon for other purposes, or rely on an external source of randomness. 


In terms of data broadcast, the overhead caused by \verloc is given by the inclusion of RTT measurements in a blockchain. Each measurement contains an RTT and an identifier of the node being measured, which together can be encoded in \SI{5}{\byte}: \SI{20}{bit} for the node id of the reference being measured (for networks of up to \num{1} million nodes) and \SI{20}{bit} for encoding the RTT with a granularity of \SI{1}{\micro\second}. Considering $R=40$ references per node (shown to be sufficient in real-world experiments) this amounts to \SI{200}{\byte} per node. The overhead scales linearly with the size of the network, \eg for a network of ten thousand nodes, the broadcast data needed to run \verloc amounts to \SI{2}{\mega\byte} for the full network. If \verloc runs once or twice a day this is a rather small overhead. A resource-efficient version of \verloc can be limited to updating information and measurements only for nodes that have newly joined or that have updated their public key, self-reported location or IP address. 

In terms of \verloc's pairwise communication overhead, each node must send \num{200} ICMP echo requests to \num{40}  references, meaning it executes \num{8000} pings, once or twice a day. This overhead remains constant as the network grows and it is in practice negligible for the nodes. Using the measurement data made publicly available, the evaluation scripts can be run locally by everyone to obtain location verification results.  Our current (non-optimized) Python scripts take less than \SI{120}{\ms} to compute the localization result for a node with $40$ references.

\section{Related Work}

\noindent\textbf{Delay-Based Geolocation.}
Different delay-based geolocation techniques~\cite{arif2010internet,eriksson2010learning,laki2011spotter,li2012ip} exist for use cases like cloud storage~\cite{gondree2013geolocation} or the identification of hidden servers~\cite{castelluccia2009geolocalization}. Such techniques overcome inaccuracies in existing databases~\cite{poese2011ip,eriksson2013understanding}, and outperform WiFi positioning systems~\cite{zandbergen2009accuracy} or GPS-based approaches~\cite{hu2007gps}. While prior studies emphasize the resilience of delay-based geolocation against \emph{simple} manipulations~\cite{muir2009internet,wang2011towards,abdou2017accurate,gill2010dude,abdou2014taxing,casado2007peering}, \verloc is the first to protect against adversaries in a decentralized setting.
Furthermore, prior approaches often rely on central authorities~\cite{gueye2006constraint,wong2007octant}. In contrast to \verloc, they depend on trusted information and do not consider targeted adversarial manipulations. Other network geolocation techniques include proxies~\cite{proxies_lie}, focus on achieving street-level granularity~\cite{wang2011towards}, or analyze the quality of the information in commercial geolocation databases~\cite{gharaibeh2017look}.

\noindent\textbf{Adversarial Localization in Other Contexts.}
Adversarial interference with location information is also relevant in other contexts. For example, prior work demonstrates privacy attacks that allow an adversary to localize and track mobile users based on public information~\cite{Shaik2015,kohls2019lost,Forsberg2007}. In the context of GPS, spoofing attacks are a persisting problem due to the unencrypted and unauthorized transmission of information~\cite{humphreys08,tippenhauer11,swaszek13b}. Several real-world incidents demonstrate the threat of such attacks~\cite{psiaki16b,russon15,bhatti14,Kremlin}.

\noindent\textbf{Distance Bounding.} 
Distance bounding protocols are two-party cryptographic protocols that enable a verifier $V$ to establish an upper bound on the physical distance to a (possibly adversarial) prover $P$~\cite{distance-bounding}. Such protocols are typically designed for bounding distances in the order of metres (or even centimetres~\cite{RF-dist-bounding2010}) and are mainly concerned with dishonest provers and man-in-the-middle adversaries that try to fake a shorter distance between $P$ and $V$~\cite{mafia-fraud}. In contrast, \verloc is a decentralized multi-party protocol that relies on redundancy across dozens of measurements with randomly selected parties to infer geolocation on a map (rather than just focusing on establishing an upper bound on the distance between two parties), such that results are accurate enough even if some measurements are fabricated by the adversary. An accuracy in the order of tens of kilometres is in most cases good enough for verifying location at country-level granularity in a global or continental network.

\section{Conclusion}

We have introduced \verloc, a decentralized protocol that uses timing probes between randomly assigned pairs of nodes to verify the geolocation of nodes in a network. \verloc outputs an estimated most likely location, a verification decision on whether the node is in the claimed geographical area (\eg, country), and a score of the confidence of \verloc on a node's localization results. Low scores are indicative of poor network conditions or of active attack. 

To configure \verloc, we first conducted an empirical study where we measured real-world propagation timings. We used this information to create a propagation model that summarizes realistic transmission times and speeds as a function of the distance between the nodes. In a series of simulation experiments, we analyzed the performance of \verloc in a non-adversarial setup and further tested its defensive capabilities in increasingly adversarial conditions. Our results show that \verloc protects against false claimed locations even when up to \SI{20}{\percent} of nodes are malicious while providing correct location estimates for the remaining benign nodes in \SI{90}{\percent} of cases. Finally, we validated \verloc with an experiment in the wild, where several thousand nodes run a \verloc prototype. The real-world results show that \verloc localizes nodes with a median error of \SI{60}{\km} and is thus suitable for verifying locations at country-level granularity. 
\subsubsection*{Acknowledgments}
We would like to thank the Nym team for enabling the VerLoc prototype. Being given the opportunity to conduct real-world measurements at this scale and in such a diverse infrastructure is unique and substantially contributed to the quality of this work. In particular, we would like to thank Dave Hrycyszyn and Jedrzej (a.k.a. ``Andrew'') Stuczynski for their implementation of the measurements. Additionally, we would like to thank Evgeny Garanin and Sergei Korolev from Nodes Guru for integrating our prototype in their system, and making the localization results of the Nym network publicly available.

This work was in part supported by the Deutsche Forschungsgemeinschaft (DFG, German Research Foundation) under Germany’s Excellence Strategy – EXC-2092 CASA – 390781972, by the Research Council KU Leuven under the grant C24/18/049, by CyberSecurity Research Flanders with reference number VR20192203, and by DARPA FA8750-19-C-0502. Any opinions, findings and conclusions or recommendations expressed in this material are those of the authors and do not necessarily reflect the views of any of the funders. 

\appendix

\section{Propagation Model}
\label{details-propagation-model}

\verloc requires a \textit{speed function} to convert measured times to distances, since a simple \textit{speed constant} does not accurately describe the relationship between the physical distance between two computers and the time it takes for data to be transmitted from one to the other via the Internet. The Internet speed function is in fact distance-dependent, with speeds being higher (and noise lower) for longer distances than for shorter ones. 

We generate a propagation model with a rather simple method that we introduced in Section~\ref{subsec:prelim:propagation} and that we describe here in more detail. We note that a better propagation model can be generated with a larger and more diverse set of samples. A more sophisticated application of the propagation model can also take into account not just the distance between two nodes, but be finer-grained and account for their actual locations since similar distances can have different speeds depending on the underlying Internet connectivity between locations. While a better propagation model can further improve \verloc's performance, we have shown in this paper that the \verloc concept works even with a simple model. 

We note that building an accurate propagation model benefits from the use of trusted landmarks at known locations. Since this only needs to be done \textit{once} before deploying \verloc and then the model is a publicly known function (that can be used not just by \verloc but by any system that wants to convert internet timings to distances), we argue that this does not introduce trusted parties in \verloc's operations. 

\subsubsection*{Setup} 

In order to build the propagation model, we measure the round-trip timings (RTTs) of transmissions between servers in \num{16} known different worldwide locations and \num{6042} relays of the Tor network. Between each pair of nodes we send \num{200} ICMP echo requests to derive the minimum RTT, the mean, and the standard deviation over all pings sent between the two servers. Along with these measured timings, we document the relays' GeoIP locations. Note that while the GeoIP location cannot serve as reliable ground truth, it provides an approximate location for the majority of nodes~\cite{gharaibeh2017look,proxies_lie}. 

Of the $200$ timing probes exchanged between a pair of nodes, we take the \textit{minimum} measured RTT as that corresponds to the \emph{least noisy} measurement sample for a connection. The same procedure is followed in \verloc whenever two nodes measure each other: they exchange $200$ probes and report the minimum measured RTT as best representative of the channel. This is because noise strictly adds (never subtracts) latency. 
\subsubsection*{Dataset} 

Within this setup, we conducted a set of 1.8 million measurements within two days, and collected two smaller sets (60k and 30k) weeks later to verify that the RTT distributions were stable. Fig.~\ref{fig:noise_overhead} documents the measurements and the fitting function, which we later use in the simulation and prototype experiments. 
We sanitized our dataset by removing nodes for which the measurements indicate they announced a false GeoIP entry. In particular, we flag measurements that would imply a propagation speed faster than light. We keep measurements that are (significantly) slower than expected, as this can be caused by network delays. The speed of light is however a hard upper bound for transmissions, and thus nodes with excessively fast measurements can be safely removed.

\section{Quality of Self-Reported Data}\label{app:self-reported}

\NEW{2}{The current prototype implementation of \verloc deployed in Nym uses self-reported locations on a city level. This can only be an approximation of the exact node location, as we refer to the city center for the latitude and longitude of a node. Furthermore, we have no access to ground truth information to check the correctness of the self-reported location. To compensate for this, we compare the city-level information provided by the node operators with the GeoIP information related to the IP addresses of the nodes. Although this does not provide us with a trusted set of locations, it provides a sanity check and a general idea of the quality of the self-reported data.}

\NEW{2}{In total, we test \num{1395} nodes of our real-world data set. This set represents the number of European nodes for which we successfully fetched GeoIP\footnote{We use the Maxmind GeoLite 2 City data base, last updated on Sep 09 2021} information. In this comparison, the median location error for all nodes, \ie, the distance between the GeoIP and the self-reported location is \SI{7.11}{\km} with a mean of \SI{364.47}{\km}. To account for the outliers, we then identify all nodes where the self-reported country and the GeoIP country contradict each other. In total we find \num{126} (\SI{9.03}{\percent}) with conflicting countries. The median location error here is \SI{1238}{\km} with a mean of \SI{3270}{\km}.} 

\NEW{2}{Finally, we revisit the results of the prototype evaluation to identify possible sources of noise leading to verification failures. To this end, we focus on two main characteristics. First, we check for inconsistencies in the self-reported and GeoIP locations. Second, we look for \verloc decisions in which a majority of \textit{slow} timing measurements influenced the confidence score. We do so for both accepted and rejected nodes. Out of \num{32768} nodes in total (tested in 64 random repetitions), \num{26993} (\SI{82.38}{\percent}) nodes were accepted and \num{5775} (\SI{17.62}{\percent}) were rejected. Out of the rejected, \num{1664} (\SI{28.81}{\percent}) had a conflicting GeoIP location, and \num{3448} (\SI{59.71}{\percent}) used measurements of which a majority (more than \SI{80}{\percent}) was unexpectedly slow. At the same time, only \num{832} (\SI{3.08}{\percent}) of all accepted nodes had a conflicting GeoIP location. Please note that these results can only represent a snapshot. Repeated runs of \verloc allow to monitor node localization and verification over time in order to better assess whether a node is truthfully reporting its location. A node that consistently fails location verification should therefore appear as more suspicious than one that occasionally fails due to slow measurements. Such longitudinal analysis approach is however out of scope for the current version of \verloc.}

\balance
{\footnotesize \bibliographystyle{plain}
\bibliography{src/references.bib}

\begin{thebibliography}{10}

\bibitem{abdou2014taxing}
AbdelRahman Abdou, Ashraf Matrawy, and Paul~C Van~Oorschot.
\newblock {Taxing the Queue: Hindering Middleboxes from Unauthorized
  Large-Scale Traffic Relaying}.
\newblock {\em IEEE Communications Letters}, 19(1):42--45, 2014.

\bibitem{abdou2017accurate}
AbdelRahman Abdou, Ashraf Matrawy, and Paul~C Van~Oorschot.
\newblock {Accurate Manipulation of Delay-Based Internet Geolocation}.
\newblock In {\em ACM Asia Conference on Computer and Communications Security},
  AsiaCCS~'17, pages 887--898, Abu Dhabi, UAE, April 2017. ACM.

\bibitem{akhoondi2012lastor}
Masoud Akhoondi, Curtis Yu, and Harsha~V. Madhyastha.
\newblock {LASTor: A Low-Latency AS-Aware Tor Client}.
\newblock In {\em IEEE Symposium on Security and Privacy}, SP~'12, pages
  476--490, San Francisco, CA, USA, May 2012. IEEE.

\bibitem{arif2010internet}
Mohammed~Jubaer Arif, Shanika Karunasekera, Santosh Kulkarni, Ajit Gunatilaka,
  and Branko Ristic.
\newblock {Internet Host Geolocation using Maximum Likelihood Estimation
  Technique}.
\newblock In {\em International Conference on Advanced Information Networking
  and Applications}, AINA~'10, pages 422--429, Perth, Australia, April 2010.
  IEEE.

\bibitem{keccak}
Guido Bertoni, Joan Daemen, Micha{\"e}l Peeters, and Gilles Van~Assche.
\newblock Keccak.
\newblock In Thomas Johansson and Phong~Q. Nguyen, editors, {\em Advances in
  Cryptology -- EUROCRYPT 2013}, pages 313--314, Berlin, Heidelberg, 2013.
  Springer Berlin Heidelberg.

\bibitem{bhatti14}
Jahshan~A. Bhatti and Todd~E. Humphreys.
\newblock {Hostile Control of Ships via False GPS Signals: Demonstration and
  Detection}.
\newblock Technical report, The University of Texas at Austin, 2014.

\bibitem{distance-bounding}
Stefan Brands and David Chaum.
\newblock Distance-bounding protocols.
\newblock In Tor Helleseth, editor, {\em Advances in Cryptology --- EUROCRYPT
  '93}, pages 344--359, Berlin, Heidelberg, 1994. Springer Berlin Heidelberg.

\bibitem{china_destroys_bitcoin}
David Canellis.
\newblock {Research: China has the power to destroy Bitcoin}, October 2018.

\bibitem{casado2007peering}
Martin Casado and Michael~J Freedman.
\newblock Peering through the shroud: The effect of edge opacity on ip-based
  client identification.
\newblock In {\em USENIX Symposium on Networked Systems Design and
  Implementation}, NSDI~'07, Cambridge, MA, USA, April 2007.

\bibitem{castelluccia2009geolocalization}
Claude Castelluccia, Mohamed~Ali Kaafar, Pere Manils, and Daniele Perito.
\newblock {Geolocalization of Proxied Services and its Application to Fast-Flux
  Hidden Servers}.
\newblock In {\em ACM SIGCOMM Conference on Internet Measurement}, IMC~'09,
  pages 184--189, Budapest, Hungary, July 2009.

\bibitem{chandrasekaran2015alidade}
Balakrishnan Chandrasekaran, Mingru Bai, Michael Schoenfield, Arthur Berger,
  Nicole Caruso, George Economou, Stephen Gilliss, Bruce Maggs, Kyle Moses,
  David Duff, et~al.
\newblock {Alidade: IP Geolocation Without Active Probing}.
\newblock {\em Department of Computer Science, Duke University, Tech. Rep.
  CS-TR-2015.001}, 2015.

\bibitem{chen2016landmark}
Jingning Chen, Fenlin Liu, Xiangyang Luo, Fan Zhao, and Guang Zhu.
\newblock {A Landmark Calibration-Based IP Geolocation Approach}.
\newblock {\em EURASIP Journal on Information Security}, 2016(1):1--11, 2016.

\bibitem{dabek2004vivaldi}
Frank Dabek, Russ Cox, Frans Kaashoek, and Robert Morris.
\newblock {Vivaldi: A Decentralized Network Coordinate System}.
\newblock {\em ACM SIGCOMM Computer Communication Review}, 34(4):15--26, 2004.

\bibitem{nym-whitepaper}
Claudia Diaz, Harry Halpin, and Aggelos Kiayias.
\newblock The {N}ym {N}etwork.
\newblock \url{https://nymtech.net/nym-whitepaper.pdf}, February 2021.

\bibitem{eriksson2010learning}
Brian Eriksson, Paul Barford, Joel Sommers, and Robert Nowak.
\newblock {A Learning-Based Approach for IP Geolocation}.
\newblock In {\em International Conference on Passive and Active Network
  Measurement}, PAM~'10, pages 171--180, Zurich, Switzerland, 2010. Springer.

\bibitem{eriksson2013understanding}
Brian Eriksson and Mark Crovella.
\newblock {Understanding Geolocation Accuracy Using Network Geometry}.
\newblock In {\em IEEE Conference on Computer Communications}, INFOCOM~'13,
  pages 75--79, Turin, Italy, April 2013. IEEE.

\bibitem{Forsberg2007}
Dan Forsberg, Huang Leping, Kashima Tsuyoshi, and Seppo Alan{\"{a}}r{\"{a}}.
\newblock {Enhancing Security and Privacy in 3GPP E-UTRAN Radio Interface}.
\newblock In {\em IEEE International Symposium on Personal, Indoor and Mobile
  Radio Communications (PIMRC)}. IEEE, 2007.

\bibitem{gharaibeh2017look}
Manaf Gharaibeh, Anant Shah, Bradley Huffaker, Han Zhang, Roya Ensafi, and
  Christos Papadopoulos.
\newblock {A Look at Router Geolocation in Public and Commercial Databases}.
\newblock In {\em ACM SIGCOMM Conference on Internet Measurement}, IMC~'17,
  pages 463--469, London, UK, November 2017. ACM.

\bibitem{gill2010dude}
Phillipa Gill, Yashar Ganjali, Bernard Wong, and David Lie.
\newblock {Dude, where’s that IP?: Circumventing Measurement-Based IP
  Geolocation}.
\newblock In {\em USENIX Security Symposium}, USENIX~'10, pages 16--16,
  Washington, DS, USA, August 2010. USENIX.

\bibitem{gondree2013geolocation}
Mark Gondree and Zachary~NJ Peterson.
\newblock {Geolocation of Data in the Cloud}.
\newblock In {\em Conference on Data and Application Security and Privacy},
  CODASPY~'13, pages 25--36, San Antonio, TX, USA, 2013. ACM.

\bibitem{gueye2006constraint}
Bamba Gueye, Artur Ziviani, Mark Crovella, and Serge Fdida.
\newblock {Constraint-Based Geolocation of Internet Hosts}.
\newblock {\em IEEE/ACM Transactions on Networking (TON)}, 14(6):1219--1232,
  2006.

\bibitem{hu2007gps}
Dexter~H Hu and Cho-Li Wang.
\newblock {GPS-Based Location Extraction and Presence Management for Mobile
  Instant Messenger}.
\newblock In {\em International Conference on Embedded and Ubiquitous
  Computing}, pages 309--320. Springer, 2007.

\bibitem{humphreys08}
Todd~E. Humphreys, Brent~M. Ledvina, Mark~L. Psiaki, Brady~W. O'Hanlon, and
  Paul~M. Kintner~Jr.
\newblock {Assessing the Spoofing Threat: Development of a Portable GPS
  Civilian Spoofer}.
\newblock In {\em International Technical Meeting of the Satellite Division of
  The Institute of Navigation}, ION GNSS~'08, pages 2314--2325, Savannah, GA,
  USA, September 2008.

\bibitem{jansen2018crowd}
Kai Jansen, Matthias Sch{\"a}fer, Daniel Moser, Vincent Lenders, Christina
  P{\"o}pper, and Jens Schmitt.
\newblock {Crowd-GPS-Sec: Leveraging Crowdsourcing to Detect and Localize GPS
  Spoofing Attacks}.
\newblock In {\em IEEE Symposium on Security and Privacy}, SP~'18, pages
  1018--1031, San Francisco, CA, USA, May 2018. IEEE.

\bibitem{katz2006towards}
Ethan Katz-Bassett, John~P. John, Arvind Krishnamurthy, David Wetherall, Thomas
  Anderson, and Yatin Chawathe.
\newblock {Towards IP Geolocation Using Delay and Topology Measurements}.
\newblock In {\em ACM SIGCOMM Conference on Internet Measurement}, IMC~'06,
  pages 71--84, Rio de Janeiro, Brazil, October 2006. ACM.

\bibitem{ouroboros}
Aggelos Kiayias, Alexander Russell, Bernardo David, and Roman Oliynykov.
\newblock Ouroboros: A provably secure proof-of-stake blockchain protocol.
\newblock In Jonathan Katz and Hovav Shacham, editors, {\em Advances in
  Cryptology -- CRYPTO 2017}, pages 357--388, Cham, 2017. Springer
  International Publishing.

\bibitem{kohls19multi}
Katharina Kohls, Kai Jansen, David Rupprecht, Thorsten Holz, and Christina
  P\"{o}pper.
\newblock {On the Challenges of Geographical Avoidance for Tor}.
\newblock In {\em Network and Distributed System Security Symposium}, NDSS~'19,
  San Diego, CA, USA, February 2019. The Internet Society.

\bibitem{kohls2019lost}
Katharina Kohls, David Rupprecht, Thorsten Holz, and Christina P{\"o}pper.
\newblock {Lost Traffic Encryption: Fingerprinting LTE/4G Traffic on Layer
  Two}.
\newblock In {\em Proceedings of the 12th Conference on Security and Privacy in
  Wireless and Mobile Networks}, pages 249--260, 2019.

\bibitem{kokoris2018omniledger}
Eleftherios Kokoris-Kogias, Philipp Jovanovic, Linus Gasser, Nicolas Gailly,
  Ewa Syta, and Bryan Ford.
\newblock {Omniledger: A Secure, Scale-Out, Decentralized Ledger via Sharding}.
\newblock In {\em IEEE Symposium on Security and Privacy}, SP~'18, pages
  583--598, San Jose, CA, USA, 2018. IEEE.

\bibitem{komosny2013can}
Dan Komosny, Milan Simek, and Ganeshan Kathiravelu.
\newblock {Can Vivaldi Help in IP Geolocation?}
\newblock 2013.

\bibitem{laki2011spotter}
S{\'a}ndor Laki, P{\'e}ter M{\'a}tray, P{\'e}ter H{\'a}ga, Tam{\'a}s
  Seb{\H{o}}k, Istv{\'a}n Csabai, and G{\'a}bor Vattay.
\newblock {Spotter: A Model Based Active Geolocation Service}.
\newblock In {\em International Conference on Computer Communications},
  INFOCOM~'11, pages 3173--3181, Shanghai, China, April 2011. IEEE.

\bibitem{byzantine-generals}
Leslie Lamport, Robert Shostak, and Marshall Pease.
\newblock The byzantine generals problem.
\newblock {\em ACM Trans. Program. Lang. Syst.}, 4(3):382–401, July 1982.

\bibitem{li2012ip}
Dan Li, Jiong Chen, Chuanxiong Guo, Yunxin Liu, Jinyu Zhang, Zhili Zhang, and
  Yongguang Zhang.
\newblock {IP-Geolocation Mapping for Moderately Connected Internet Regions}.
\newblock {\em Transactions on Parallel and Distributed Systems},
  24(2):381--391, 2012.

\bibitem{Li2017DeTorPA}
Zhihao Li, Stephen Herwig, and Dave Levin.
\newblock {DeTor: Provably Avoiding Geographic Regions in Tor}.
\newblock In {\em USENIX Security Symposium}, USENIX~'17, pages 343--359,
  Vancouver, BC, Canada, August 2017. USENIX Association.

\bibitem{mafia-fraud}
Aikaterini Mitrokotsa, Cristina Onete, and Serge Vaudenay.
\newblock Mafia fraud attack against the rČ distance-bounding protocol.
\newblock In {\em 2012 IEEE International Conference on RFID-Technologies and
  Applications (RFID-TA)}, pages 74--79, 2012.

\bibitem{muir2009internet}
James~A Muir and Paul C~Van Oorschot.
\newblock {Internet Geolocation: Evasion and Counterevasion}.
\newblock {\em Acm computing surveys (csur)}, 42(1):1--23, 2009.

\bibitem{poese2011ip}
Ingmar Poese, Steve Uhlig, Mohamed~Ali Kaafar, Benoit Donnet, and Bamba Gueye.
\newblock {IP Geolocation Databases: Unreliable?}
\newblock {\em ACM SIGCOMM Computer Communication Review}, 41(2):53--56, 2011.

\bibitem{psiaki16b}
Mark~L. Psiaki and Todd~E. Humphreys.
\newblock {Attackers can spoof navigation signals without our knowledge. Here's
  how to fight back GPS lies}.
\newblock {\em IEEE Spectrum}, 53(8):26--53, August 2016.

\bibitem{RF-dist-bounding2010}
Kasper~Bonne Rasmussen and Srdjan Capkun.
\newblock Realization of rf distance bounding.
\newblock In {\em Proceedings of the 19th USENIX Security Symposium}, pages 389
  -- 401, Washington, DC, 2010. USENIX Association.
\newblock 19th USENIX Security Symposium 2010; Conference Location: Washington,
  DC, USA; Conference Date: August 11-13, 2010.

\bibitem{claps}
Florentin Rochet, Ryan Wails, Aaron Johnson, Prateek Mittal, and Olivier
  Pereira.
\newblock Claps: Client-location-aware path selection in tor.
\newblock In {\em Proceedings of the 2020 ACM SIGSAC Conference on Computer and
  Communications Security}, CCS '20, page 17–34, New York, NY, USA, 2020.
  Association for Computing Machinery.

\bibitem{russon15}
Mary-Ann Russon.
\newblock {Wondering how to hack a military drone? It's all on Google}, May
  2015.

\bibitem{Kremlin}
Clare Sebastian.
\newblock {Getting lost near the Kremlin? Russia could be 'GPS spoofing'},
  December 2016.

\bibitem{Shaik2015}
Altaf Shaik, Ravishankar Borgaonkar, N.~Asokan, Valtteri Niemi, and Jean-Pierre
  Seifert.
\newblock {Practical Attacks Against Privacy and Availability in 4G/LTE Mobile
  Communication Systems}.
\newblock In {\em Network and Distributed System Security Symposium}. The
  Internet Society, 2016.

\bibitem{shavitt2011geolocation}
Yuval Shavitt and Noa Zilberman.
\newblock {A Geolocation Databases Study}.
\newblock {\em IEEE Journal on Selected Areas in Communications},
  29(10):2044--2056, 2011.

\bibitem{swaszek13b}
Peter~F. Swaszek and Richard~J. Hartnett.
\newblock {Spoof Detection Using Multiple COTS Receivers in Safety Critical
  Applications}.
\newblock In {\em International Technical Meeting of The Satellite Division of
  the Institute of Navigation}, ION GNSS+~'13, pages 2921--2930, Nashville, TN,
  USA, September 2013.

\bibitem{random-beacon}
E.~{Syta}, P.~{Jovanovic}, E.~K. {Kogias}, N.~{Gailly}, L.~{Gasser},
  I.~{Khoffi}, M.~J. {Fischer}, and B.~{Ford}.
\newblock {Scalable Bias-Resistant Distributed Randomness}.
\newblock In {\em IEEE Symposium on Security and Privacy}, SP~'17, pages
  444--460, San Jose, CA, USA, 2017. IEEE.

\bibitem{tippenhauer11}
Nils~Ole Tippenhauer, Christina P\"{o}pper, Kasper~Bonne Rasmussen, and Srdjan
  \v{C}apkun.
\newblock {On the Requirements for Successful GPS Spoofing Attacks}.
\newblock In {\em ACM Conference on Computer and Communications Security},
  CCS~'11, pages 75--86, Chicago, IL, USA, October 2011. ACM.

\bibitem{SoK-decentralization}
Carmela Troncoso, George Danezis, Marios Isaakidis, and Harry Halpin.
\newblock Systematizing decentralization and privacy: Lessons from 15 years of
  research and deployments.
\newblock {\em CoRR}, abs/1704.08065, 2017.

\bibitem{1498470}
S.~\v{C}apkun and J.~P. Hubaux.
\newblock {Secure Positioning of Wireless Devices with Application to Sensor
  Networks}.
\newblock In {\em IEEE Conference on Computer Communications}, INFOCOM~'05,
  pages 1917--1928, Miami, FL, USA, March 2005. IEEE.

\bibitem{wang2011towards}
Yong Wang, Daniel Burgener, Marcel Flores, Aleksandar Kuzmanovic, and Cheng
  Huang.
\newblock {Towards Street-Level Client-Independent IP Geolocation}.
\newblock In {\em USENIX Symposium on Networked Systems Design and
  Implementation}, NSDI~'11, pages 27--27, Boston, MA, USA, March 2011. USENIX
  Association.

\bibitem{proxies_lie}
Zachary Weinberg, Shinyoung Cho, Vyas Sekar, and Phillipa Gill.
\newblock {How to Catch when Proxies Lie: Verifying the Physical Locations of
  Network Proxies with Active Geolocation}.
\newblock In {\em ACM SIGCOMM Conference on Internet Measurement}, IMC~'18,
  Boston, MA, USA, October 2018. ACM.

\bibitem{wong2007octant}
Bernard Wong, Ivan Stoyanov, and Emin~G{\"u}n Sirer.
\newblock {Octant: A Comprehensive Framework for the Geolocalization of
  Internet Hosts}.
\newblock In {\em USENIX Symposium on Networked Systems Design and
  Implementation}, NSDI~'07, pages 23--23, Santa Clara, CA, USA, June 2007.
  USENIX Association.

\bibitem{zandbergen2009accuracy}
Paul~A. Zandbergen.
\newblock {Accuracy of iPhone Locations: A Comparison of Assisted GPS, WiFi and
  Cellular Positioning}.
\newblock {\em Transactions in GIS}, 13:5--25, 2009.

\bibitem{coindesk}
Wolfie Zhao.
\newblock {Top Bitcoin Mining Pools See 15\% Hashrate Drop Amid Continuous
  Rainstorms in China}, August 2020.

\end{thebibliography}

\end{document}
